\documentclass[12pt,english]{article}
\usepackage[applemac]{inputenc}
\usepackage[letterpaper,top=1in,bottom=1in]{geometry}
\usepackage{array}
\usepackage{booktabs}
\usepackage{multirow}
\usepackage{amsmath}
\usepackage{amssymb}
\usepackage{setspace}
\usepackage{subcaption}
\makeatletter

\usepackage[parfill]{parskip}
\usepackage{graphicx}
\usepackage[hang]{footmisc}
\usepackage{booktabs}\usepackage[english]{babel}
\usepackage{graphicx}

\usepackage[round]{natbib}

\usepackage{xcolor}
\usepackage{epsfig}
\usepackage{tcolorbox}
\usepackage{authblk}

\usepackage{epstopdf}

\begin{document}

\begin{titlepage}

\title{Reducing residential emissions: carbon pricing vs. subsidizing retrofits}

\author[1,2]{Alkis Blanz}
\author[3]{Beatriz Gaitan}

\affil[1]{\footnotesize{University of Leipzig, Faculty of Economics and Management Sciences, Grimmaische Str. 12, 04109, Leipzig, Germany}}
\affil[2]{Mercator Research Institute on Global Commons and Climate Change (MCC), Torgauer Str. 12-15, 10829, Berlin, Germany}
\affil[3]{Potsdam Institute for Climate Impact Research, Telegrafenberg, 14473, Potsdam, Germany}

\date{\today}

\maketitle
\begin{abstract}
In this paper, we compare different mitigation policies when housing investments are irreversible. We use a general equilibrium model with non-homothetic preferences and an elaborate setup of the residential housing and energy production sector. In the first-best transition, the energy demand plays only a secondary role. However, this changes when optimal carbon taxes are not available. While providing subsidies for retrofits results in the lowest direct costs for households, it ultimately leads to the highest aggregate costs and proves to be an ineffective way to decarbonize the economy. In the second-best context, a phased-in carbon price outperforms the subsidy-based transition.
\end{abstract}
\thanks{\footnotesize{We are very grateful to Matthias Kalkuhl, Kai Lessmann, Ottmar Edenhofer for useful discussion and comments. This project has received funding from the European Union's Horizon Europe research and innovation programme under grant agreement No 101056891  - CAPABLE - ClimAte Policy AcceptaBiLity Economic framework. Views and opinions expressed are however those of the author(s) only and do not necessarily reflect those of the European Union. Neither the European Union nor the granting authority can be held responsible for them.}}
\end{titlepage}

\restoregeometry
\section{Introduction}
Reducing the carbon emissions in the housing sector requires large upfront investments. Investment needs include retrofitting the existing housing stock and constructing new, more energy-efficient buildings to meet housing demand \citep{iea2021netzero}. However, vast evidence indicates that current investment levels are insufficient, increasing the possibility of a carbon lock-in \citep{ipcc2022buildings}. In this situation, investments are inconsistent with climate targets, leaving only expensive reductions through stranded assets as an option for emission reductions. Preventing a carbon lock-in in residential housing calls for ambitious mitigation policy that can stimulate and coordinate various types of investment. 

In the context of housing investments, the optimal instrument choice for mitigation is unclear. Housing-related investments are different. Investments in housing are irreversible \citep{miles2009irreversibility} and therefore cannot be converted after construction. Furthermore, climate policy in housing is challenging from a political economy perspective, as shelter is a basic need for humans. Implementing ambitious climate policy can be difficult to implement. Imposing high carbon taxes on households increase housing costs and utility bills, which may increase poverty. Furthermore, if the direct costs for households are too high, households may not be able to invest sufficiently in energy efficiency and the expansion of renewable energy. 

This paper addresses this gap and compares different mitigation policies when housing investments are irreversible. We build a multisector general equilibrium model with a special focus on the different investment decisions affecting the decarbonization of the residential housing sector. The model combines an elaborate setup of the residential housing market alongside energy markets. We separate investments to reduce the energy intensity of the existing housing stock from investments in the construction of new, more energy-efficient buildings. Furthermore, we allow for investments in fossil and non-polluting capital used in energy production. We compare different carbon pricing scenarios with transitions that rely on housing-related investment subsidies and calibrate the model to the case of Germany. Thereby, we isolate the impact of policy constraints for the green transition.

When policymakers face no constraints on the availability of mitigation policies, the conventional wisdom holds and a uniform carbon price is the cost-effective instrument for the green transition \citep{pigou1932effect,goulder2008instrument,FISCHER2008}. In the first-best scenario, energy demand plays only a secondary role. However, this changes when optimal carbon taxes are not available. When it is not possible to directly increase fossil resource prices, climate policy has to rely on reductions in energy demand to reduce fossil fuel consumption in housing. Due to the irreversibility constraint, the only option is to excessively subsidize investments in energy efficiency, making the subsidy-based transition very costly from a social perspective.

This paper builds on several strands of the literature. First, we add to the literature that analyzes optimal timing and allocation of abatement investments across sectors. \cite{vogt2018starting} find that it is optimal to start with significant short-term abatement investment because we cannot switch overnight to carbon-free technologies. It matters in which sectors the short-term effort abatement investment happens \citep{vogt2014marginal,lecocq1998decision,vogt2014marginal}. We extend this literature by explicitly considering housing investments related to the energy demand of the housing stock. We are able to show that housing investments differ from investments in industrial capital. In contrast to industrial capital, investing in housing capital directly increases utility at the cost of having a higher energy demand. The role of energy efficiency investments is to solely reduce these direct costs of housing investments. One contribution is to show how climate policy optimally affects the different investment incentives over time.

Furthermore, we relate to the large literature that focuses on the energy efficiency gap \citep{JAFFE1994,Allcott2012,Gerarden2017}, the difference between optimal and actual energy use. While the existing literature considers both behavioral failures and market failures as potential causes for inefficient energy use, it focuses largely on static frameworks. We complement this literature by analyzing the intersection within a dynamic, general-equilibrium setting. Thereby, we are able to consider the interaction of energy demand in housing with climate policy and macroeconomic activity. We contribute to this literature by analyzing the energy efficiency gap in a dynamic setting within the context of reducing residential carbon emissions.

The closest paper to ours is \citet{ROZENBERG2020}. The authors compare the impact of different mitigation policies in a multisector general equilibrium model with irreversible investment and a climate constraint. We consider our paper to be complementary to their analysis. \citet{ROZENBERG2020} discuss a potential trade-off between the political feasibility and cost-effectiveness of mitigation policies based on the premature retirement of fossil capital. Instead, we consider housing costs, including energy bills, as the main source of transition costs for households. Thereby, we expand their insights for energy production by focusing on the case of residential housing.

The paper is structured as follows. Section 2 describes the model environment. Section 3 summarizes the calibration, while section 4 presents the main analysis. Finally, section 5 concludes. 
\newpage{}

\section{Model Environment}
This section presents the dynamic multi-sector general equilibrium model used for analyzing the transition towards a low carbon economy. The model combines an elaborate setup of the housing and energy sectors, as well as an optimizing government as in \citet{kalkuhl2012learning}. The dynamic setting allows for studying the interaction between sectors and actors. The decisions of households determine how energy-intensive their housing demand is. Energy producers decide on how carbon-intensive energy production is, and the government sets policy instruments optimally to achieve a climate target. 

\subsection{Households}

The economy is populated by $n$ households who have preferences of type:

\begin{equation}
\max\sum_{t=0}^T\frac{1}{\left(1+\rho\right)^{t}}\frac{\left(c_{t}^{\phi}\left(h_{t}-\bar{h} \right)^{1-\phi}\right)^{1-\eta}}{1-\eta},
\end{equation}

where $\rho$ is the pure rate of time preference, $\eta^{-1}>0$ is the elasticity of intertemporal substitution and $\phi$ is a share parameter.\footnote{Depending on the application, we either use the pure time preference rate $\rho$ or the discount factor $\beta$. Recall that they are directly related: $\rho=(1-\beta)/\beta$.} Households derive utility from final consumption $c_t$ and from housing services $h_t$. Housing services are subject to a subsistence level $\bar{h}$ that captures the minimal need for shelter for humans. Thus, preferences are non-homothetic, as in \citep{geary1950note,stone1954linear}.

Households have access to a wide portfolio
of assets. These assets include industry capital used in the production of the final good and electricity, as well as different assets related to the production of housing services. As the focus lies on aggregate dynamics, we assume that all households are property owners and produce their own housing services. Households produce housing services according to:
\begin{equation}
h_{t}=land_{t}^{a_{h}}(\bar{k}+k_{t}^H)^{1-a_{h}}
\end{equation}

The production of housing services requires $land_t$ and housing capital $\bar{k}+k_t^H$. We distinguish between old housing capital ($\bar{k}$) with a high energy intensity and capital $k^H_t$ with a low energy efficiency. The total amount of land is assumed to be fixed. Thus, by expanding the housing stock, households can consume more housing services and thereby increase their utility. However, the construction of new buildings affects the energy demand of housing $ene_t$. The energy demand of the house covers both the appliances of the house and heating.
Housing-related energy demand is described by:
\begin{equation}
ene_t =  \kappa_o (k^E)  \bar{k} + \kappa_N k_{t}^H
\end{equation}

Depending on the energy efficiency of new buildings $\kappa_N>0$, expanding the housing stock will increase the total energy demand of the housing stock. Additionally, the total energy demand depends on the energy demand of the old housing stock $\bar{k}$. Especially in urban areas, old buildings are maintained so that they remain in use.
The parameter $\kappa_N>0$ and the function $\kappa_o$, explained below, describe the energy efficiency of each housing stock. We assume that households pay maintenance costs on the existing housing stock to overcome its depreciation so that the old stock remains constant. The existence of maintenance costs is a common feature of housing models \citep{piazzesi2016housing,henderson1983model,li2007life}. Essentially, this simplification captures the fact that part of the housing stock is of such high value to society that it will not be demolished even if the energy intensity of these buildings is high. Instead, we allow for investments in energy efficiency capital $k^E$. The stock of efficiency capital captures all kinds of retrofits, such as improvements to the insulation of walls, windows, and the roof. Thus, investments in efficiency capital reduce the energy intensity $\kappa_o$ of the existing stock, which is:

\begin{equation}
\kappa_o(k^E) =\frac{\bar{\kappa}}{k_t^E} + \kappa_N 
\end{equation}

where $\bar{\kappa}$ is a positive constant. Thus, there are limits to the improvements in energy efficiency of the existing housing stock. In detail, the function form implies that in the limit, the energy demand of old buildings cannot exceed the energy efficiency of new buildings $\kappa_N$. Investments in housing capital and efficiency capital determine the total energy demand of the housing stock. Increasing the production of housing services by expanding the housing stock decreases the overall energy intensity, as new buildings are considered to be more energy efficient but increases the overall energy demand. In contrast, efficiency capital does not increase the production of housing services but decreases the energy demand of the housing stock.
 
The energy demand for housing determines utility bills and, thus, housing costs. Depending on their investment choices into housing and efficiency capital, utility bills will be higher or lower. The energy demand related to housing can be satisfied by using electricity and a fossil resource. We assume that electricity $e_{t}$
and the fossil resource $res_{t}$ are imperfect substitutes for households. 
\begin{equation}
ene_{t}=\left(a_{ene}e_t^{\frac{\sigma_{ene}-1}{\sigma_{ene}}}+(1-a_{ene})res_{t}^{\frac{\sigma_{ene}-1}{\sigma_{ene}}}\right)^{\frac{\sigma_{ene}}{\sigma_{ene}-1}}
\end{equation}

The elasticity of substitution between electricity and the fossil
resource is denoted $\sigma_{ene}>0$, and $a_{ene}$ and $(1 -a_{ene})$
are their respective share parameters. Households pay for electricity and for the use of the fossil resource. The use of the fossil resource may be subject to a carbon price $p_{HC,t}$. Therefore, the expenditure side
of the households consists of consumption expenditures, investment
$inv^i_t$ in the portfolio of industry capital, land, and the two stocks of housing-related capital, and
the energy costs attached to them.\\

Land, industry capital (final good capital ($k_t^Y$) and fossil and renewable energy capital ($k_t^F$ and $k_t^N$)), housing, and efficiency capital evolve according to

\begin{subequations}
\begin{align}
land_{t+1}= & i_{t}^{Land}+land_{t}\\
k_{t+1}^{Y}= & i_{t}^{Y}+\left(1-\delta_{Y}\right)k_{t}^{Y}\\
k_{t+1}^{F}= & i_{t}^{F}+\left(1-\delta_{F}\right)k_{t}^{F}\\
k_{t+1}^{N}= & i_{t}^{N}+\left(1-\delta_{H}\right)k_{t}^{N}\\
k_{t+1}^{H}= & i_{t}^{H}+\left(1-\delta_{F}\right)k_{t}^{H}\\
k_{t+1}^{E}= & i_{t}^{E}+\left(1-\delta_{N}\right)k_{t}^{E}
\end{align}
\end{subequations}

where $i^j_t$ denotes the respective investment, and $\delta_{j}$ is the rate of capital depreciation. In a dynamic
general equilibrium model, agents must be indifferent between
investing in industry capital, housing capital, and energy efficiency capital. Thus, households will consider not only energy and land prices but also the interest rate when making investment decisions. This is reflected in
the non-arbitrage conditions, which can be found in the appendix. 

Housing-related investments differ from investments in industry capital. Investments in housing capital and energy efficiency capital are irreversible as in \citet{arrow1970optimal}. Once households invest in housing, capital cannot be transformed into consumption or industry capital. The irreversibility is summarized by:
\begin{equation}
   inv_{t}^{H} \geq 0
\end{equation}
\begin{equation}
   inv_{t}^{E} \geq 0
\end{equation}
\subsection{Production}

On the production side of the economy, we assume the production of
a final good and energy. The final good
is used for consumption and for investments. The production
technology of the final good producer is described by:

\begin{equation}
Y_t=\left(a Z_t^{\frac{\sigma-1}{\sigma}}+\left(1-a\right)E_{Y,t}^{\frac{\sigma-1}{\sigma}}\right)^{\frac{\sigma}{\sigma-1}}
\end{equation}

For production, the final good producer relies on two intermediate
inputs. The final good producer combines final electricity $E_{Y,t}$
from the electricity producer and a capital-labor composite. The capital-labor
composite equals: 
\begin{equation}
Z_t=K_{Y,t}^{a_{Z}}\left(A_{Y,t}L_{t}\right)^{1-a_Z} 
\end{equation}

where $K_{Y,t}$ and $L_{t}$, respectively, denote the capital and labor employed in the production of $Y_t$, and $A_{Y,t}$ is exogenous labor augmenting technological change. We follow \cite{Barrage2023} so that $A_{Y,t}$ evolves according to

\begin{equation}
A_{Y,t}=A_{Y,t-1}/\left( 1-a_{Y}e^{-b_{Y}\left( t-1\right) }\right) 
\end{equation}

where $a_Y$ and $b_Y$ are positive constants. The production side of the economy allows for an elaborate description
of the electricity sector that resembles in many aspects the approach employed in \cite{kalkuhl2012learning}. The final electricity producer demands electricity from fossil ($E_{F,t}^{d}$) and non-polluting renewable sources
($E_{N,t}^{d}$). The production function of the final energy
firm is: 
\begin{equation}
E_{t}=\left(a_{E}\left(E_{F,t}^{d}\right)^{\frac{\sigma_{E}-1}{\sigma_{E}}}+\left(1-a_{E}\right)\left(E_{N,t}^{d}\right)^{\frac{\sigma_{E}-1}{\sigma_{E}}}\right)^{\frac{\sigma_{E}}{\sigma_{E}-1}} 
\end{equation}

where $a_{E}$ is a share parameter and $\sigma_{E}$ is
the elasticity of substitution between $E_{F,t}^{d}$ and $E_{N,t}^{d}$.
The role of the final energy firm in our setting is that of an electricity
provider. The firm does not produce the electricity itself by operating
wind farms and power plants, but buys the electricity from fossil
and renewable electricity producers. Thus, the profits of the final electricity firm are described by: 
\begin{equation}
\pi_{E,t}=p_{E,t}E_{t}-p_{F,t}E_{F,t}^{d}-p_{N,t}E_{N,t}^{d}   
\end{equation}

where $p_{E,t}$ denotes the final price of energy and $p_{F,t}$ and $p_{N,t}$, respectively, denote the prices of fossil and renewable energy.

The fossil energy producer uses capital and fossil resources to produce electricity. The production function of the polluting electricity
firm is: 
\begin{equation}
E_{F,t}=\left(a_{F}K_{F,t}^{\frac{\sigma_{F}-1}{\sigma_{F}}}+\left(1-a_{F}\right)Res_{F,t}^{\frac{\sigma_{F}-1}{\sigma_{F}}}\right)^{\frac{\sigma_{F}}{\sigma_{F}-1}}
\end{equation}
where $a_{F}$ is a are share parameters, and $\sigma_{F}$ is
the elasticity of substitution between$K_{F,t}$ and $Res_{F,t}$. Based on the setup in \citet{kalkuhl2012learning}, we assume that the use of a fossil resource is the source of carbon emissions and not the use of a dirty capital stock. We abstract from fossil resource production within the economy. Instead, we assume a small-open economy that imports the fossil resource at a price
$p_{R,t}$. Climate policy can affect the full price of the fossil resource by imposing a carbon price
$p_{C,t}$. Thus, the profits of the polluting fossil electricity
firm are described by:

\begin{equation}
\pi_{F,t}=p_{F,t}E_{F,t}-R^F_tK_{F,t}-(p_{R,t}+p_{C,t})Res_{F,t}   
\end{equation}

In contrast to the polluting firm, the renewable energy producer does
not rely on fossil resources to produce electricity. Producing non-polluting
electricity is a function of capital.
The production technology and profits of the non-polluting electricity
firm are described by: 
\begin{equation}
E_{N,t}=A_{N,t}K_{N,t}    
\end{equation}

\begin{equation}
\pi_{N,t}=p_{N,t}E_{N,t}-R^N_tK_{N,t}    
\end{equation}

$A_{N,t}$ is exogenous technological change that evolves according to

\begin{equation}
A_{N,t+1}=A_{N,t}/\left(1-g_{N}\right).    
\end{equation}

Given a rental rate of capital $r_t$, the cost of producing renewable energy declines at the exogenous rate $g_{N}$. This specification resembles the decrease in cost of the backstop technology employed in \cite{Barrage2023}. 

\subsection{Government}

The primary focus of the government in our setup is
climate policy. As in \cite{kalkuhl2012learning}, the government is an agent that, acting as a Stackelberg leader of its own economy, chooses policy instruments optimally under a given set of constraints. We assume that the government has a fixed emission
target in the form of a carbon budget and a set of policy tools to achieve
the target. For expositional simplicity, we assume that carbon emissions are directly associated with the use of the fossil resource both in housing and energy production. By assuming a common emission target, we fix the level
of ambition across scenarios and allow for a comparison of the effectiveness of different policy instruments. The climate target is given by:

\begin{equation}
    \overline{M} = \sum_{t=0}^{T} (Res_{F,t}+nres_{t})
\end{equation}

The main climate policy tool is a
carbon price that taxes the use of the fossil resource. However, we
allow for differentiated carbon prices in housing and industry. Carbon pricing is the sole revenue source for the government in our setting. The budget constraint of the government in the baseline analysis is:

\begin{equation}
\Gamma_{t}+n\tau_{INVE,t}inv_{t}^{E}
=p_{C,t}Res_{F,t}+np_{HC,t}res_{t}    
\end{equation}

where $\tau_{INVE,t}$ is a subsidy to energy efficiency capital investment. Depending on the set of policies considered, the revenues from carbon pricing can be used either to pay transfers to households $\Gamma_{t}$ or to subsidize investments into efficiency capital.\footnote{Theoretically, it could also be an option to subsidize investments into housing capital, but as it is never optimal in our setting, we omit it from the budget constraint for expositional simplicity.}

\subsection{Market clearing}
This section briefly presents the market clearing conditions in our setup. Capital market clearing equals
\begin{equation}
K_{Y,t}=nk_{t}^{Y}\text{, }K_{F,t}=nk_{t}^{F}\text{, }K_{N,t}=nk_{t}^{N}\text{;}\label{eq:MCCK}
\end{equation}

Market clearing in electricity equals
\begin{equation}
E_{Y,t}+ne_{t}=E_{t};
\end{equation}
The market clearing in fossil energy is:
\begin{equation}
E_{F,t}^{d}=E_{F,t};
\end{equation}
Similarly, the market clearing in renewable energy equals
\begin{equation}
E_{N,t}^{d}=E_{N,t};
\end{equation}
Finally, the balance of payments is balanced if
\begin{align}
Y_{t}= & n\left(c_{t}+i_{t}^{Y}+i_{t}^{F}+i_{t}^{N}+i_{t}^{H}+i_{t}^{E}+\delta_{H}\overline{k}\right)\label{BP}\\
 & +p_{R,t}\left(n\times res_{t}+Res_{F,t}\right)\nonumber 
\end{align}

\begin{table}[!tph]
\caption{Calibration}

\begin{tabular}{lllll}
& Parameter & description & value &  \\ \hline\hline
& $a$ & share of capital-labor composite $Z$ & \multicolumn{1}{r}{$0.95$} & 
\\ 
& $\sigma $ & elasticity of substitution $Z-E_{Y}$ & \multicolumn{1}{r}{$0.40
$} &  \\ \cline{2-4}
& $a_{Z}$ & share of capital $K_{Y}$ in $Z$ & \multicolumn{1}{r}{$0.35$} & 
\\ 
& $A_{Y,0}$ & labor productivity at $t=0$ & \multicolumn{1}{r}{$1.00$} &  \\ 
\cline{2-4}
& $a_{E}$ & share of fossil energy $E_{F}$ & \multicolumn{1}{r}{$0.57$} & 
\\ 
& $\sigma _{E}$ & elasticity of substitution $E_{F}-E_{N}$ & 
\multicolumn{1}{r}{$4.70$} &  \\ \cline{2-4}
& $a_{F}$ & share of capital $K_{F}$ & \multicolumn{1}{r}{$0.80$} &  \\ 
& $\sigma _{F}$ & elasticity of substitution $K_{F}-Res_{F}$ & 
\multicolumn{1}{r}{$0.20$} &  \\ \cline{2-4}
& $A_{N,0}$ & capital $K_{N}$ productivity at $t=0$ & \multicolumn{1}{r}{$%
0.69$} &  \\ \cline{2-4}
& $\phi $ & share of final good consumption & \multicolumn{1}{r}{$0.80$} & 
\\ 
& $\bar{h}$ & housing services minimum consumption & \multicolumn{1}{r}{$1.62
$} &  \\ 
& $a_{H}$ & share of land in housing services & \multicolumn{1}{r}{$0.35$} & 
\\ \cline{2-4}
& $a_{ene}$ & share of electricity $e$ in housing energy & 
\multicolumn{1}{r}{$0.14$} &  \\ 
& $\sigma _{ene}$ & elasticity of substitution $e-res$ & \multicolumn{1}{r}{$%
0.89$} &  \\ 
& $\kappa _{N}$ & new housing capital energy intensity & \multicolumn{1}{r}{$%
0.02$} &  \\ 
& $\bar{\kappa}$ & parameter in $\kappa _{o}\left( k_{t}^{E}\right) $
function & \multicolumn{1}{r}{$0.005$} &  \\ \cline{2-4}
& $\eta ^{-1}$ & elasticity of intertemporal substitution & 
\multicolumn{1}{r}{$1.00$} &  \\ 
& $\rho $ & consumers rate of time preference & \multicolumn{1}{r}{$0.02$} & 
\\ \cline{2-4}
& $\delta _{Y}=\delta _{F}=\delta _{N}$ & industry depreciation rate of
capital & \multicolumn{1}{r}{$0.10$} &  \\ 
& $\delta _{H}$ & depreciation rate of housing capital & \multicolumn{1}{r}{$%
0.02$} &  \\ 
& $\delta _{E}$ & depreciation rate of energy efficiency capital & 
\multicolumn{1}{r}{$0.03$} &  \\ 
& $Land=n\ast land$ & Aggregate land endowment & \multicolumn{1}{r}{$501.23$}
&  \\ 
& $k_{0}^{E}/\overline{k}$ & Ratio $k_{0}^{E}$ to $\overline{k}$ & $0.08$ & 
\\ \cline{2-4}
& $a_{Y}$ & constant in labor augmenting technological change & $0.0859$ & 
\\ 
& $b_{Y}$ & growth parameter in labor augmenting technological change & $%
0.0072$ &  \\ 
& $g_{N}$ & growth parameter of technological change & \multicolumn{1}{r}{$%
0.01$} &  \\ \hline\hline
\end{tabular}

\end{table}

\section{Calibration}

We calibrate the model to the German economy and indicate the parameter
values used throughout the simulations in Table 1.

\subsection{Consumers}

For the share $\phi$ of final good consumption, we use German input-output data for the year 2014 from the World Input-Output Database (WIOD, refer to \cite{Timmer2015}). Let $x_{c}$ denote aggregate household expenditures except for real estate activities and energy-related expenditures (mining and quarrying, manufacture of coke and refined petroleum products, and electricity, gas, steam, and air conditioning supply) from the WIOD. Let $x_{h}$ denote the expenditure on real estate activities, and $x_{ene}$ denote energy-related expenditures from the WIOD. Aggregate expenditure, thus, equals $x\equiv{x_{c}+x_{h}+x_{ene}}.$ The share $x_{c}/\left( x_{c}+x_{h}\right)$ equals $0.81$, we round this number and set $\phi =0.80$. The housing services composite is similar to \cite{Combes2021} who find a share parameter for capital equal to $0.65$ using French data. For lack of a German estimate we set $\left( 1-a_{h}\right)$ equal to $0.65$ so that the share of land $
a_{h}$ equals $0.35$.

For the housing energy parameters $a_{ene}$ and $\sigma_{ene}$, we focus on heating space and hot water, which together account for about $80\%$ of total residential energy use in Germany (refer to \cite{IEA2021}). For the housing fossil energy share parameter $\left( 1-a_{ene}\right)$, we employ the estimates of direct residential fossil energy (coal, gas, and oil) from the Institute for Housing and Environment (IHE, refer to \cite{IHE2012}). To
account for the large share of fossil energy in district heating, we add $70\%$ of residential district heating energy
(as reported by the IHE) to the direct fossil energy use and generate a total
fossil energy use.\footnote{More than $70\%$ of district heating is produced using fossil resources according to the \cite{BMWK2021}.} The ratio of total fossil
energy use to total energy use
equals 0.854. Since more than $70\%$ of district heating
is produced using fossil resources, we round up $0.854$ and set $
\left(1-a_{ene}\right)$ equal to $0.86$, so that the share of electricity $\left( a_{ene}\right) $ equals $0.14$. 

Let $s_{e}$ denote the
household expenditure share of electricity out of total household energy expenditure. Let $\varepsilon _{res}$ denote the households' fossil resource partial own price elasticity of demand. With a CES function $\varepsilon
_{res}=-s_{e}\sigma _{ene}$ (cf. \cite{Allen1938} p. 373) where $\sigma _{ene}$ is the elasticity of substitution between electricity and fossil resources by households. Since natural gas is the largest source of energy for space and water heating in Germany, we use Nilsen et al. (2012) estimate of the short-run price elasticity of demand for natural gas of German households. Their estimate equals $-0.131$.  We set $s_{e}=\left(1-0.854\right)$, setting $\varepsilon _{res}=-0.131$ and using $\varepsilon_{res}=-0.131=-s_{e}\sigma _{ene}=-\left( 1-0.854\right) \sigma _{ene}$
implies $\sigma _{ene}=0.89,$ which is the value we use for the elasticity of
substitution between electricity and fossil resources used by households.\footnote{An elasticity smaller than 1 indicates that fossil fuels and electricity are complements rather than substitutes in housing. While historically this seems intuitive, investments in district heating or heat pumps may affect this elasticity. Due to the focus on energy efficiency, we leave this aspect for future research.}

We set the rate of time preference $\rho $ equal to $0.02$, a value often used
elsewhere (see \cite{Barro2003}). In view of \cite{Guvenen2006}
findings, we set the inverse of the elasticity of intertemporal substitution
$\eta $ equal to 1.

\subsection{Depreciation and production}

\cite{Davis2015} survey the macroeconomic housing
literature and find that the estimates of the depreciation rate of
housing structures are in the range of 0.01 to 0.03. Based on this, we set
the depreciation rate of housing capital $\left( \delta _{H}\right) $ equal
to $0.02$. For the depreciation rate of energy efficiency capital, $\left(
\delta _{E}\right)$, we use the depreciation rate of electricity, oil, and
wood heaters of $0.03$ used by \cite{Nesbakken2001} who analyzes the energy consumption of the heating of space for the case of Norway. For the rate of industry capital
depreciation ($\delta _{K}$), we employ the value of 0.10 used by \cite{Kiyotaki2011}, a value that is also consistent with the long-run rate of capital depreciation in manufacturing found in \cite{Albonico2014}. The separation of housing capital leads to larger estimates for the depreciation of manufacturing capital than the values typically employed for aggregate capital.

Regarding the elasticity of substitution between the capital-labor composite and electricity of the final good ($\sigma$), we employ estimates of \cite{Koesler2015}. Their estimate of all industries is equal to 0.38. We
round this number and set $\sigma =0.4$. The share of the capital-labor composite of the final good is proxied by making a sectoral aggregation of the input-output matrix of the WIOD. We aggregate the input-output matrix into four sectors, namely i)
fossil resources, ii) electricity, gas, steam, and air conditioning supply,
iii) real estate activities, and iv) all the remaining sectors which we think of as the final good. We set the share of the capital-labor composite in the production of the final good $\left( a\right)$ equal to the share of value-added in energy expenditures (i and ii) plus value added. This leads
to $a=0.95$.

Regarding the capital share
of the capital-labor composite $\left( a_{Z}\right)$, we use \cite{Valentinyi2008} estimate. They find that the omission of intermediate inputs leads to biased values if capital and labor shares are directly estimated from input-output tables. They solve this by
calculating the amount of capital and labor embodied in intermediate inputs and impute this to sectoral capital and labor shares. No estimates for Germany are available, but the US and German economies are sufficiently similar such that we use their estimate. Valentinyi and Herrendorf estimate capital shares of agriculture, manufactured consumption, services, equipment investment, and construction; and consider various aggregations of these five sectors. Their aggregation of agriculture, manufactured consumption, and services leads to a capital share of $0.35$. We set the capital share equal to this value $\left(a_{Z}=0.35\right)$

For the elasticity of substitution between fossil and clean energy, $\left(\sigma_{E}\right)$ and the respective share parameters of the energy composite, we use the German estimates of \cite{Stoeckl2020}. We use the average of their current and future elasticity of substitution estimates
(the latter takes into account capacity constraints), the average equals $4.7$, we thus set $\sigma _{E}=4.7$. We take a similar average from Stoeckl and Zerrahn in the case of the share of clean energy $\left( 1-a_{E}\right)$ leading to $0.43$, and thus set the share of fossil energy $\left(a_{E}\right)$ equal to $0.57$.

In the case of the elasticity of substitution between capital and the fossil
resource $\left(\sigma _{F}\right)$ and the capital share parameters $\left(a_{F}\right)$ in the production of fossil energy, we use the values used by \cite{kalkuhl2012learning} and thus set $\sigma _{F}=0.15$, and $a_{F}=0.8$.

The parameters $a_Y$ and $b_Y$ in Table 1 reproduce the values employed by \cite{Barrage2023}, and $g_N$ decreases the cost of clean energy by 1

The rest of the parameters $\bar{h},$ $\bar{\kappa},$ $\kappa _{N}$, $A_{N,0},$ and the $Land,$ and the ratio $k^{E}/\bar{k}$ of respective capital endowments are set so that a steady-state solution, absent of policy and technological progress, satisfies the following properties.

According to the German Federal Statistical Office (\cite{Destatis2022}), housing costs (including energy) were on average
$23.3\%$ of disposable income of households. We solve the steady state of the model
without policy so that the expenditure on housing services and energy on
disposable income equals $0.233$.

In Germany during the year 2019, a $70m^{2}$ well-insulated apartment
building with gas heating had an average heating cost of $485$ euros,
instead, the average heating costs of a $70m^{2}$ apartment in a
badly insulated building amounted to $1,030$ euros (see \cite{CLEW2020}). We use this information and solve the steady state of the model without policy so that the ratio of the energy efficiency of old housing capital to that of new housing capital, that is $\left( \bar{\kappa}/k_{E}+\kappa _{N}\right)
/\kappa _{N}$ equals $1,030/485$.

According to \cite{BMWK2022}, the share of renewables in the German electricity sector was $41\%$ in the year 2021. We use this value and solve the steady state of the model without policy so that the ratio of clean energy to clean and fossil energy $E_{N}/(E_{N}+E_{F})$ equals 0.41.

The emissions generated in the German electricity and building sectors (residential, commercial, and military), respectively, amounted to $247$ and $115$ million tonnes of CO$_{2}$ equivalents in 2021 (refer to \cite{Statista2022a}). Of the building emissions, $76\%$ were residential emissions; in other words, residential emissions amounted to about $115*0.76=87.4$ million tonnes of $CO_{2}$ equivalents (see \cite{Statista2022b}). We solve the steady state of the model without policy so that the ratio of residential fossil consumption to residential plus electricity fossil consumption $n* res/(n*res+Res_F)$ equals 87.4/(87.4+247) =0.26

The Institute for Housing and Environment (see \cite{IHE2012}) provides the area of single and multifamily housing units in Germany. They divide housing into different groups depending on when they were built. We aggregate the single and multifamily houses that were constructed until 1978, the year in which the Ordinance on Thermal Insulation went into effect. We consider those houses constructed until 1978 to be the old housing capital stock with a higher energy intensity and those constructed after to be of low energy intensity. The share structures constructed before 1978 equals $66\%$. We solve the steady state without policy so that the ratio $\bar{k}/\left( \bar{k}+k^{H}\right) =0.66.$

In the steady-state solution without policy, we set the price of land equal to one so that the aggregate land endowment can be determined. The values of $\bar{h}$, $\bar{\kappa}$, $\kappa _{N}$, $A_{N,0}$, and the $Land$, and the ratio $k^{E}/\bar{k}$ are indicated in Table 1.

\section{Results}
The transition to a carbon-free economy depends on ambitious climate policy. In the absence of climate policy, the economy continuously relies on the use of fossil resources for both housing and energy production. The aim of this section is to study various transitions towards a carbon-free economy. All transitions are optimal given the respective constraints, as all agents, including the government, behave optimally. However, in all scenarios, the government has access to distinct sets of policy instruments. Specifically, we analyze the first-best transition without limitations on the availability of different policy instruments and compare it to several second-best settings. The main restriction we consider is on carbon pricing in the housing sector. In detail, we compare the first-best transitions to transitions where carbon pricing in housing is phased-in or is entirely unavailable.

When analyzing the first-best transition, we rely additionally on analytical derivations from the model setup. Thereby, we are able to isolate how climate policy can incentivize the investment shifts necessary for decarbonizing the economy. Additionally, we can confirm that the conventional wisdom holds in our setup. The optimal instrument is a uniform carbon price in all sectors of the economy. Interestingly, the irreversibility constraint does not affect optimal instrument choice. Instead, the constraint matters for the housing-related energy demand across the transition path. To analyze the optimal instrument choice and the implications of the irreversibility constraint, we rely on numerical simulations of the model.

\subsection{Optimal Transition}

The analysis begins by comparing the optimal transition to the no-policy benchmark. In all scenarios, the economy grows due to exogenous technological progress. The increase in economic output and household prosperity lead to a higher demand for energy and housing. Thereby, the demand for the fossil resource increases as well. The fossil resource enters both the supply, as well as the demand side of the economy. Fossil energy producers use the resource alongside capital to produce energy. Additionally, households use the fossil resource alongside electricity to satisfy the energy demand of their housing demand. In the laissez faire, these decisions are not affected by climate policy.

The transition requires changing the price of using the fossil resources, as the externality linked to the use of the resource is not accounted for. However, apart from the resource use, internalizing these external costs may also influence investment behavior of households. In our setup, households invest in both industry and housing capital. Industry capital covers capital that relates to energy and final good production. Housing and efficiency capital are relevant for producing housing services. These two types of investment exhibit fundamental differences and react differently to climate policy. Through analytical derivations of the model, we can explore this nexus in more detail.

At the core of our general equilibrium model is a rich investment portfolio. As these investments are interdependent, households choose investments such that they are indifferent between the different investment options. Conceptually, investing in capital for energy production, whether it be fossil or non-polluting, provides an alternative to investing in physical capital. All three are forms of industrial capital. Households choose consumption and investment in the laissez-faire equilibrium such that the values of physical capital, as well as clean and fossil capital, are equalized:

\begin{equation}
    \psi_t=\psi_t^C=\psi_t^F=\lambda_t
\end{equation}

where $\psi_t$ is the value of investments in physical capital, $i \in (F,N)$ are the shadow values of investments in fossil and non-polluting capital, and $\lambda_t$ is the shadow value of income. In the end, fossil and clean capital are simply another way to invest in physical capital. As perfect capital mobility exists, the return of these assets is determined essentially by the Euler equation:

\begin{align}
R_{t+1}= & \frac{1}{\lambda_{t+1}}\left[\frac{\psi_{t}}{\beta}-\left(1-\delta_{Y}\right)\psi_{t+1}\right]
= \frac{u_{c_{t}}}{\beta u_{c_{t+1}}}-\left(1-\delta_{Y}\right)
\end{align}

Without climate policy, the marginal productivity of fossil and clean capital, as well as their returns, are equal:\footnote{ The only possible difference in returns are differences in depreciation rates. }
\begin{equation}
R_{t}=R^F_t=R^N_t
\end{equation}

Climate policy affects investment in energy-related capital stocks indirectly through its effect on incentives for energy producers. Without a carbon price, the fossil energy producer chooses fossil resource use by setting its value marginal productivity equal to the exogenous price $p_{R,t}$. However, the price of fossil energy is too low from a societal perspective since it is not high enough to satisfy the emissions budget. With a carbon price, the demand for non-polluting capital increases. Through a relative price change, electricity producers rely less on fossil energy, which increases the incentive to move capital to renewable energy production. The change in marginal productivity of fossil and non-polluting capital affects the return on investments and, thus investment behavior of households. Optimal climate policy shifts investments from fossil capital to non-polluting capital.

Housing investments differ from investments in industrial capital. An increase in housing capital leads to a higher level of housing services, making housing investments a direct source of utility rather than increasing income in the future. The implicit return of investing in housing is utility. Non-homothetic preferences determine the increase in utility from additional housing consumption and, thereby, the return from investing in housing capital. Housing services are subject to a subsistence level. Depending on household income levels, the increase in housing capital increases utility differently. For example, an additional unit of housing consumption increases utility more strongly at low-income levels as households spend a relatively larger share of their income on housing. Since an increase in housing capital increases utility more strongly at low-income levels, it becomes more attractive to invest in housing capital. Thus, investing in housing capital becomes less attractive over time as households become wealthier.

However, investments in housing capital have direct costs. Increasing the stock of housing capital also increases the aggregate energy demand of the housing stock. This energy demand is met by households using electricity and the fossil resources. Consequently, an increase in housing capital results in higher energy expenditures for households. This increase in energy expenditures lowers the return of an increase in housing capital (cf. non-arbitrage conditions in the appendix). The size of this increase depends on the energy efficiency of newly constructed buildings. With a higher energy efficiency of new buildings, the impact of these direct costs decreases.

The rich investment portfolio in our setup allows for an alternative that lowers the direct costs of housing capital. By retrofitting the existing housing stock, households can directly lower the housing-related energy demand. This decrease in energy demand can offset the direct costs of housing investments.  Increasing the energy efficiency capital stock lowers the direct costs of housing capital by lowering the overall energy expenditures of households. Because the housing stock becomes less energy-intensive, households need less electricity and fossil resources to meet the energy demand. Expanding the housing stock with new, energy-efficient buildings affects less the energy expenditure of households. 

Another difference from industrial capital, in our setting, is that housing investments are irreversible. Irreversibility refers to the fact that once capital has been invested in a certain activity, it cannot be transformed into consumption goods \citep{arrow1970optimal}. Housing is reversible only at very high costs, which is equivalent to irreversible investment \citep{miles2009irreversibility}. Therefore, households need to anticipate that investing in housing-related capital means that it cannot be converted back into consumption or industry capital. The irreversibility constraints have a direct effect on the shadow value of housing and efficiency capital. The shadow values of housing capital and efficiency capital can be expressed as:
\begin{equation}
    \psi_t^H=\lambda_t + \phi_t^H,
\end{equation}
and
\begin{equation}
    \psi_t^E=\lambda_t + \phi_t^H,
\end{equation}
where $\phi_t^i$,  $i\in\left\{ H,E\right\} $ is the multiplier of the respective irreversibility constraint of housing and efficiency capital (for more details, refer to the appendix).

If the irreversibility constraint is binding, then $\phi_t^i >0$ which creates a wedge between the shadow value of housing-related capital and industry capital. However, when the irreversibility constraints are not binding, the shadow values of all types of capital are equal. To determine whether these constraints are binding, it is useful to analyze how climate policy affects the incentives to invest in both types of housing capital.

In contrast to industry capital, climate policy directly affects the incentives for investing in housing-related capital. In the absence of climate policy, the price of fossil resources for heating does not account for external costs associated with the burning of fossil fuels. This implies that the private and social value of energy demand in housing differ (In appendix \ref{app_analytical}, we provide expressions for the shadow value of energy demand $\nu_t^{ene}$ in both cases. Without climate policy, the shadow value of energy demand is lower. Thereby, a marginal increase in housing capital is less costly in terms of increased energy demand:
\begin{align}
\frac{\partial h^{ene}}{\partial k_{t+1}^{H}}= & \frac{1}{\nu_{t+1}^{ene}}\left[\chi_{t+1}\frac{\partial h^{k}}{\partial k_{t+1}^{H}}+\left(1-\delta_{H}\right)\psi_{t+1}^{H}-\frac{\psi_{t}^{H}}{\beta}\right],
\end{align}

where $\chi_t$ equals the marginal utility of housing services and $\delta_H$ is the depreciation rate of housing capital. The direct costs of investing in housing capital are too low from a social perspective. In this situation, households reasonably invest more in housing capital, as this increases utility. The rise increase in energy costs due to the higher energy demand is less significant due to the low price of both electricity and fossil fuels. Therefore, compared to the scenario with optimal climate policy, households over-invest in housing capital.

From a societal perspective, the low direct costs of housing investments make it less attractive to invest in reducing the energy demand. Investing in efficiency capital solely lowers energy expenditures by lowering the energy demand of the existing housing stock. Without climate policy, the marginal benefit of decreasing the energy demand is smaller: 
\begin{align}
\frac{\partial h^{ene}}{\partial k_{t+1}^{E}}= & \frac{1}{\nu_{t+1}^{ene}}\left[\left(1-\delta_{E}\right)\psi_{t+1}^{E}-\frac{\psi_{t}^{E}}{\beta}\right],
\end{align}
where $\delta_E$ is the depreciation rate of efficiency capital. In the laissez faire, the return to investing in retrofitting the existing stock is too low. Therefore, the stock of efficiency capital is lower than in the optimal policy case. Climate policy affects housing investments by increasing the shadow value of energy demand. From this, we can derive first insights for the irreversibility constraints. First, as the stock of efficiency capital is always lower in the no-policy scenario, the irreversibility constraint on efficiency capital will not be a barrier to the transition to a carbon-free economy. It will always be optimal to expand the stock of efficiency capital. Therefore, in the subsequent analysis, we only focus on the irreversibility constraint on housing capital investments.

The irreversibility constraint on housing investments may be binding, as the stock of housing capital is higher without climate policy. In principle, it may be optimal to have a lower energy demand once the carbon budget is introduced, as energy is more expensive. Decreasing the housing capital stock partially mitigates this problem by reducing the energy demand and, thus, energy expenditures. However, housing capital differs from industrial capital. Lowering the housing capital stock decreases housing services and, consequently, the level of utility. The extent of this loss of utility depends on non-homothetic preferences. Furthermore, there is an alternative available in the situation. By Investing in efficiency capital, households effectively use a backstop technology. When the irreversibility constraint binds, households can invest in efficiency capital and directly decrease the pressure of the constraint. In our setting, only households only want to decrease sharply housing capital, if the energy demand is too high. 

Essentially, the question is what households do when they realize that the housing stock is too large. The first option is to decrease the energy demand by decreasing the housing capital at the expense of utility. If this is the cheapest option, then the irreversibility constraint is a barrier. The second option is to forego present consumption in favor of investing in efficiency capital. Investing in retrofitting the existing stock may be a suitable alternative to running down the housing capital stock. Theoretically, investing in industrial capital through the expansion of capital in non-polluting electricity lowers the costs of electricity. In our general equilibrium model, these options interact. It is challenging to determine analytically if the constraint is binding. Thus, we explore whether the constraint is binding through the numerical simulation of the model.

The last part of the analytical derivations focuses on optimal climate policy. Given the differentiated impact on investments, it may be optimal to use differentiated carbon prices for households and the industry. Furthermore, decarbonizing electricity production may facilitate the decarbonization of the housing sector if electricity becomes cheaper. This leads to the question of whether it is optimal to implement different carbon prices for housing and energy production.

We find that the conventional wisdom, namely the least-cost theorem \citep{Baumol1971} holds: the optimal carbon price is equal for households and industry. The intuition is as follows. In our setting, emissions are associated with the use of the fossil resource. During the transition, the substitution possibilities vary on the production and the household side. While in the industry, the expansion of renewable energy is the main option, for housing, electricity is the primary alternative. At the same time, reducing the energy demand in final goods production and housing may be necessary. In order to incentivize these changes, a shift in relative prices is necessary. A uniform carbon price optimally achieves this shift. The optimal carbon tax rate is given by:
\begin{equation}
\tau_t =\tau_t^Y = \frac{\mu_t^R }{u_{c_t} p_{R,t}}=\tau_t^h
\end{equation}

In turn, the optimal carbon price equals $p_{C,t}=\tau_t  p_{R,t}$.  The level of the carbon tax is determined by the shadow value of the carbon budget $\mu_t^R$ divided by the marginal utility of consumption $u_{c_t}$. This is standard in the literature (i.e., \citet{barrage2018careful,kalsbach2021pricing}). The only difference is that in our setting, it additionally depends on the (exogenous) price of the resource $p_{R,t}$. 
The transition to a carbon-free economy features a gradual decline in the use of the fossil resource both in energy production and in housing. A gradual increase in carbon prices ensures that the decline in fossil resources is optimal.\footnote{In the appendix, we show that this increase is constant and equal to the pure rate of time preference.} The transition relies on a shift in investment behavior, which we analyze in more depth in the numerical simulation.

\begin{figure}
    \centering
    \includegraphics[scale=0.65]{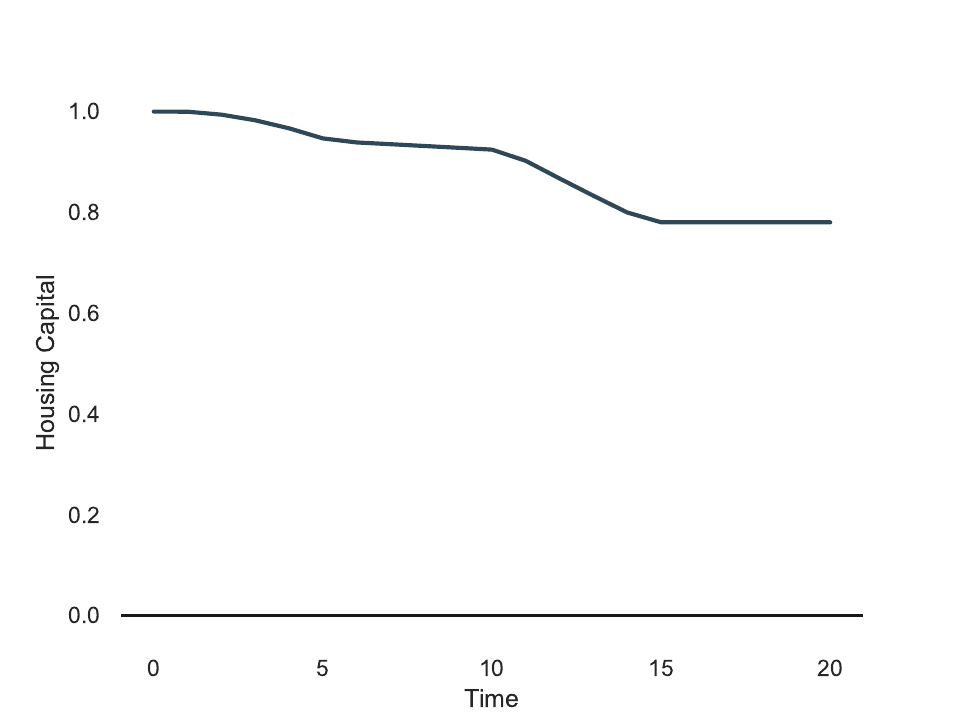}
    \caption{Housing capital stock relative to the no-policy benchmark}
    \label{figkH}
\end{figure}

\begin{figure}
    \centering
    \includegraphics[scale=0.6]{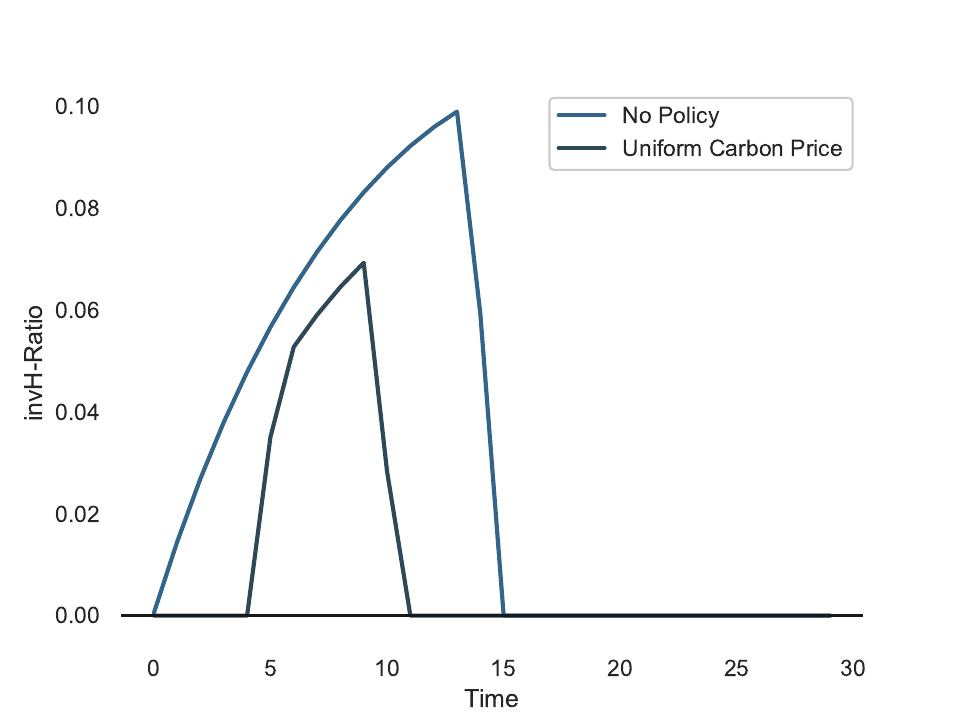}
    \caption{Housing capital investments in the optimal policy scenario and the no-policy baseline. Investments are expressed as a ratio over industry capital investments.}
    \label{invH}
\end{figure}

\begin{figure}[h!]
  \centering
  \begin{subfigure}{.5\textwidth}
    \centering
    \includegraphics[width=\linewidth]{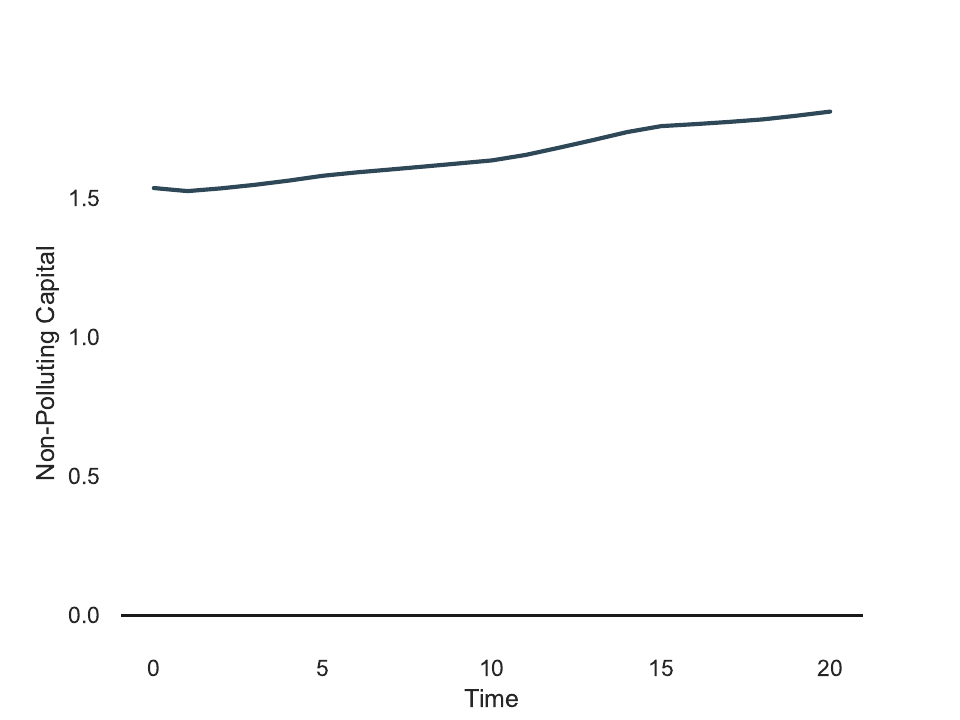}
  \end{subfigure}%
  \begin{subfigure}{.5\textwidth}
    \centering
    \includegraphics[width=\linewidth]{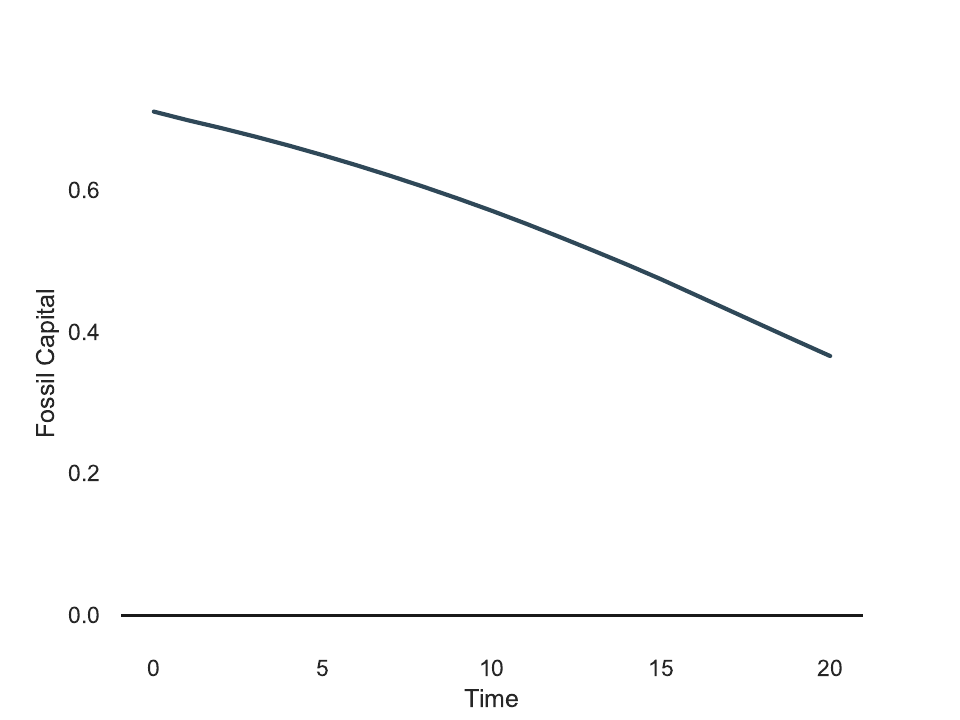}
  \end{subfigure}
 \caption{Evolution of the different capital stocks in the optimal climate policy scenario in deviation from the no-policy benchmark.}
    \label{fbcapital}
\end{figure}

Within the numerical simulation, the climate target is to reduce carbon emissions by 50 percent within three decades. In our setting, this is equivalent to reducing fossil resource use by the same amount. Figure \ref{fbcapital} describes capital accumulation dynamics throughout the transition for various forms of industry capital and for efficiency capital. The appendix includes graphs with further variables such as resource use, carbon prices, and renewable energy. The variables are presented as deviations from the no-policy path. Since there is exogenous technological progress, we compare the first-best transition to the outcome, where there is no an emission budget constraint but only technological progress during the same period. 

During the first-best transition, energy production steadily decreases the fossil energy share and relies increasingly on renewable energy. By imposing a carbon tax on the fossil resource, climate policy increases the costs of the fossil energy producers and subsequently the price of fossil energy. As a result, the final energy producer substitutes fossil energy with renewable energy. This is reflected by a strong increase in non-polluting capital, accompanied by a decline in fossil capital. After a decade, the fossil capital stock is already half of what it is in the no-policy scenario. At the same time, the stock of non-polluting capital is 60 percent greater than the respective stock in the laissez-faire path.

The higher energy costs also influence the production of the final good. As the green transition requires taxing a previously cheaper good, energy, it results in a loss in output. Higher energy prices alter energy demand in final good production, which affects output. While there is a decline in output during the transition, the decline is not driven by overall lower investments into industry capital.  Interestingly, industry capital investments do not change significantly. In fact, during the optimal transition, industry capital investments closely resemble those from the no-policy path.
Next, we turn to the investment dynamics in the housing sector. Retrofitting becomes more attractive through climate policy, resulting in an increase in the energy efficiency capital stock ($k^E_t$). After two decades, the efficiency capital stock is 50 percent higher than in the no-policy case. The increase in efficiency capital occurs in two phases. Initially, it is optimal to rapidly increase the stock of efficiency capital, while the carbon price remains low. Afterwards, the stock of efficiency capital increases only gradually until it reaches a plateau. Overall, the energy demand of the housing stock decreases by 30 percent over this period. However, energy demand does not only depend on changes in the stock of efficiency capital. Additionally, investments in housing capital affect the aggregate energy demand of the housing stock. Figure \ref{figkH} shows the evolution of the housing capital stock relative to the no policy benchmark. In the optimal climate policy scenario, the housing stock is 20 percent lower in the long run. To analyze the role of the irreversibility constraint in more detail, it is useful to look at investment rather than the stock of housing capital.

Figure \ref{invH} describes the dynamics of housing investments for both the no-policy and optimal-policy paths. Due to the irreversibility constraint, we focus on investments themselves rather than the housing capital stock. Housing capital investments are expressed as a ratio over industry capital investments. Without climate policy, housing investments steadily increase until they reach their peak. Afterwards, households do not invest further into housing capital. With advancements in technology, households become richer and eventually reach a point where additional housing service do not provide sufficient utility to outweigh increased energy expenditures. The total expansion of housing capital is significant without climate policy. At their peak, housing investments are one-tenth of physical capital investments. 

In contrast, with optimal carbon pricing in place, the total expansion of the housing stock is limited. When considering the external costs of fossil fuel use, the direct costs of expanding the housing stock make it less attractive to invest in housing capital. Consequently, housing investments peak earlier and at a lower level. Since the increase in energy demand is more costly, households reach faster the point where an increase in utility from additional housing services is not worth it anymore. Thus, it is optimal to stop earlier investing further in housing capital. 

Additionally, housing investments start later than in the no-policy case and only after a few periods. This indicates that the irreversibility constraint is binding. It is optimal to reduce sharply residential energy demand in the first period. Households achieve this in two ways. First, households initially run-down part of the housing stock. Since the irreversibility constraint prevents negative investments, the only option is depreciation. Furthermore, the presence of the irreversibility constraint triggers the large investments in efficiency capital. Thus, in the first-best transition, it is optimal to reduce energy demand in the short term by both reducing the housing capital stock and sharply increasing the amount of retrofits.
\begin{figure}[h!]
  \centering
  \begin{subfigure}{.5\textwidth}
    \centering
    \includegraphics[width=\linewidth]{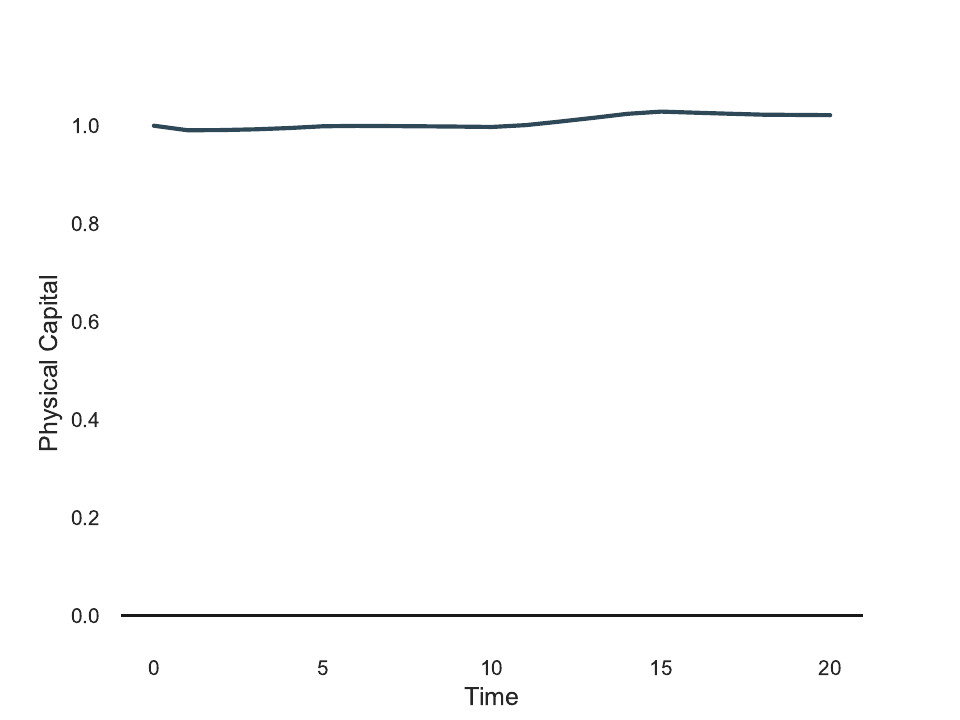}
  \end{subfigure}%
  \begin{subfigure}{.5\textwidth}
    \centering
    \includegraphics[width=\linewidth]{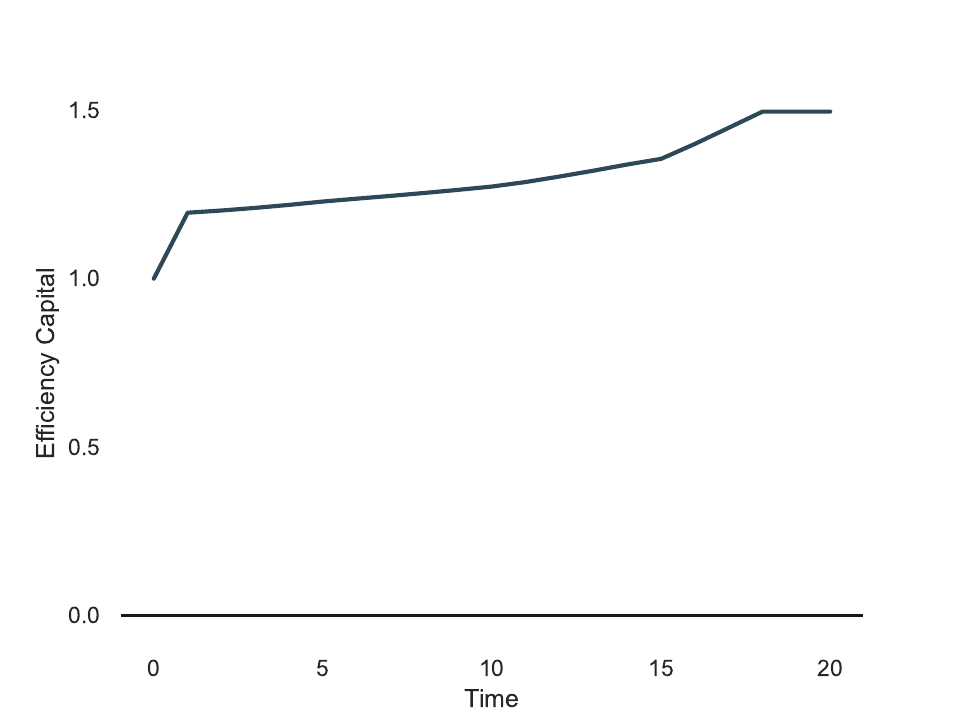}
  \end{subfigure}
\end{figure}

\subsection{Second-best Transition}
Implementing optimal carbon pricing schemes may not be feasible. Imposing high carbon prices on households may cause resistance and thus may lack political support. Households may have misperceptions about the effectiveness or fairness of carbon taxes \citep{Douenne2022}. In addition, depending on their political background, policymakers may prefer subsidies or regulatory policies to the introduction of taxes. Given that high carbon prices are needed to decarbonize the housing sector, we compare the first-best transition with transitions, where carbon prices for households can only slowly be introduced or are not available at all. Thus, we consider a phased-in carbon price scenario and one scenario, where only subsidies for efficiency capital are used. Note that these transitions are optimal in the sense that all actors, including the government, behave optimally. Thus, the government sets taxes and subsidies optimally given the constraints it faces. 

In all scenarios, we do not affect climate policy on the production side. The aim is to consider different policy scenarios for the housing sector while leaving energy production unaffected. We achieve this by separating the carbon budget. The carbon budget for the industry is set to the optimal burden in the first best. Since we do not impose any restrictions on climate policy in energy production, the resulting carbon price path for the industry is identical to the first best. This allows us to isolate the effects of different restrictions on climate policy on the household side. This approach is inspired by climate policy on the European level, where separate emission trading systems are being established for firms and households. In the present analysis, our focus is on climate policy on the household side. In detail, we consider the scenario of phased-in carbon prices as in \cite{ROZENBERG2020}, as well as investment subsidies for efficiency capital.\footnote{In \ref{app_fig} appendix, we include a graphical comparison of investment patterns for housing capital and efficiency capital for all scenarios. The scenarios include the no-policy benchmark, first-best climate policy, and the described second-best settings.}

\begin{figure}
    \centering
    \includegraphics[scale=0.8]{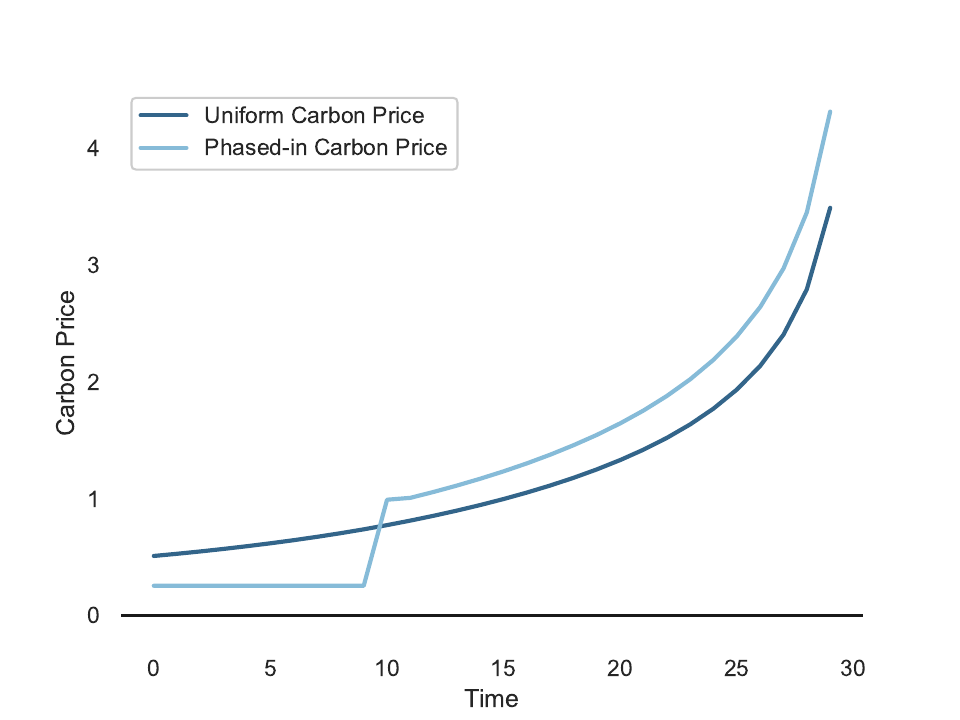}
    \caption{Optimal versus phased-in carbon prices. Variables are expressed as a ratio over fossil resource price.}
    \label{fig:cp}
\end{figure}

The phased-in carbon price scenario focuses on short-term restrictions on the level of carbon prices. Figure \ref{fig:cp} compares the phased-in carbon price for households to the optimal carbon price. The underlying intuition is that a slower implementation of carbon pricing will lead to lower immediate costs for households and thus increase political support compared to a higher carbon price regime. This approach is common for the introduction of carbon price schemes. Examples include  Phased-in carbon prices reduce emissions more slowly in the short run, because we restrict them in the first decade. During this period, the carbon price is fixed to a lower level than the first-best carbon price. In detail, due to the phased-in carbon price, the price of fossil resource use increase by 25 percent. Afterwards, the government optimally sets the carbon price. 

Additionally, we consider the case where carbon pricing is not available at all for the household side. In this case, the alternative is investment subsidies. In principle, the government can subsidize both the construction of new energy-efficient buildings and the retrofit of the existing stock. However, subsidizing investments in housing capital is never optimal in our setting. While new construction increases the energy efficiency of the overall housing stock, it also increases the energy demand. Since a higher level of energy demand makes the transition more costly, it is not useful from a climate policy perspective to subsidize housing capital investments. Thus, we solely consider subsidies to energy efficiency capital in the case where carbon pricing for housing is not available.

From the first-best transition, we understand that the irreversibility constraint on housing capital interacts with the necessity to reduce energy demand. One solution is to lower the energy demand of the existing stock. To measure the difference in energy efficiency between old and new buildings, we use the energy efficiency gap.\footnote{The energy efficiency gap is defined as the energy efficiency ratio of old to new buildings $EEG = \frac{\kappa_N + \frac{\bar{\kappa}}{k_t^E} }{\kappa_N}$. We normalize it by the initial level of the ratio.} When the efficiency gap reaches zero, the stock of old buildings has the same energy intensity as new buildings. Thereby, the energy efficiency gap determines the limits to investments in efficiency capital. Particularly in subsidy-based transitions, the energy efficiency gap may act as a barrier for climate policy.

\begin{figure}
    \centering
    \includegraphics[scale=0.8]{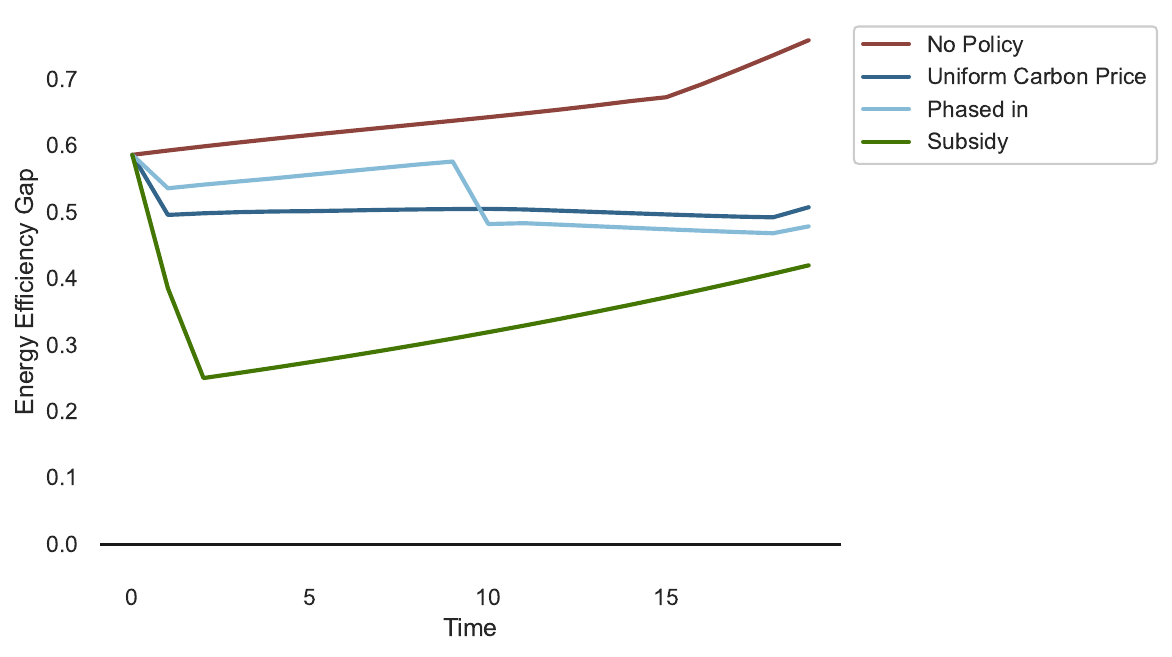}
    \caption{Evolution of the Energy Efficiency Gap in the first-best, second-best scenarios and the no-policy benchmark}
    \label{EEG}
\end{figure}

Figure \ref{EEG} describes the evolution of the energy efficiency gap in housing for both the first-best and 
second-best transitions, as well as the no-policy benchmark. Without climate policy, the energy efficiency gap grows over time. Energy costs decrease over time, reducing the return to investing in energy efficiency and, thereby, the level of efficiency capital. In contrast, all scenarios with climate policy result in a significantly smaller energy efficiency gap. However, energy demand plays a fundamentally different role in carbon price and subsidy-lead transitions. In transitions with carbon pricing, energy demand plays a secondary role. While energy demand is relevant for the irreversibility constraint, the reduction of fossil resource use is managed by affecting the price directly. If carbon pricing is not available, lowering the energy demand becomes the primary tool to reduce fossil fuel consumption in housing. Thus, investments in efficiency capital are essential. Notably, even in this scenario, it is never optimal to close the efficiency gap completely. We will analyze next the second-best transitions in greater depth.

We start with the case of phased-in carbon prices. As the carbon budget is identical to the first best, carbon prices must be higher once the policy constraint on the level of carbon taxes is no longer binding. Thus, the phased-in carbon price scenario focuses on a shift in the level of ambition during the transition. In exchange for a lower level of ambition in earlier periods, the level of ambition increases more afterwards to meet the climate target. The lower carbon price in the initial period leads to relatively lower energy expenditures for households. While the direct burden on households is lower, the lower carbon price weakens the price signal to increase energy efficiency investments.

When carbon prices are phased in, the investment pattern for efficiency capital differs from the first best. While in the first best, it is optimal to increase investments in energy efficiency sharply in the initial period, the initial expansion in the phased-in scenario is much smaller. In contrast to the first best, there is a second wave of investment in energy efficiency in the last period of the constrained carbon prices. Specifically, households invest a second time in efficiency capital. Notably, the magnitude of the increase in investment is almost identical to the initial expansion. The underlying intuition is that it is optimal for households to split investments into efficiency capital, as the initial increase addresses the comparatively lower carbon price. The second increase takes advantage of the fact that the burden is still low from the initial carbon price scheme, but due to technological progress, households are richer than in the initial period. 

In contrast, when the government uses subsidies for efficiency capital, investment dynamics differ both qualitatively and quantitatively. First, the price of the fossil resource cannot be directly influenced. Reducing the use of fossil resource requires a reduction in energy demand. By reducing the energy demand of the housing stock, households need less electricity and less fossil resource to meet their energy demand. Consequently, very large investments in efficiency capital are needed to decrease the energy demand of the housing stock sufficiently. It is optimal to invest early and strongly in efficiency capital. In detail, the investments are five times higher than in the first-best scenario. After the large investments in the early periods, energy efficiency investments are zero thereafter.

Next, we turn to investments in housing capital. Compared to the first-best investment pattern, investments in housing capital are moved forward. While similar in magnitude to the first best, investment in housing capital occurs after the investment hike in efficiency capital, but as long as carbon prices are constrained. Due to the lower carbon price, the irreversibility constraint on the housing stock is less binding in the early periods. In contrast, housing capital investment is larger when climate policy relies on efficiency capital subsidies. Intuitively, when the energy demand of existing stock decreases due to the large expansion of efficiency capital, the direct cost of investing in housing capital is smaller. Hence, it is more attractive to expand the stock of housing capital to receive additional utility. 

\begin{table}[]
    \centering
\begin{tabular}{lccc}\hline \hline  
&  Uniform &   Phased-in &   Subsidy\\ \hline 
Welfare &    -0.0443 &    -0.0449  &    -0.0565 \\ 
Energy Costs   &    0.597 &    0.616  &    -0.243 \\ 
Housing Costs (net)  &    0.022 &    0.023  &    0.01 \\ 
Housing Costs    &    0.0553 &    0.0587  &    -0.001 \\ 
Disp. Income    &    0.0057 &    0.006  &    -0.018 \\ 
Transfers & 0.015 & 0.015 & -0.003 \\
Output   &    -0.03 &    -0.03  &    -0.034 \\ 
Energy Prod.   &    0.058 &    0.058  &    0.044 \\ 

\hline \hline
\end{tabular}
    \caption{Comparison of uniform carbon price, phased-in carbon price and the investment subsidy for efficiency capital. Changes in cumulative welfare, energy costs, net housing costs, housing costs (incl. energy), disposable income, output and energy production relative to the laissez-faire scenario. Household transfers are expressed as percent of income (net of transfers). All variables are discounted at the internal discount rate}
    \label{tab:welfare}
\end{table}

The irreversibility constraint is directly related to investments in housing capital. In the first best, households want to decrease housing capital stock during the initial period once they realize that the stock is excessively large, resulting in inefficiently high energy expenditures. Ideally, households immediately want to reduce the size of the housing stock, but the irreversibility constraint is preventing this. In the phased-in carbon price scenario, the irreversibility constraint is less problematic. Due to the lower carbon prices in the earlier periods, the need to reduce the energy demand in the first period is lower. In contrast, the irreversibility constraint is a key barrier in the transition that relies on subsidies. The underlying reason is that this transition has to use energy demand reduction as the primary tool to reduce fossil fuel use. Due to the irreversibility constraint, reducing the housing stock is no option, making investments in efficiency capital necessary. However, the large amounts of efficiency-related investments crowd out other types of investments, such as investments in industry capital.

Apart from the investment dynamics, the different policies affect welfare. Table \ref{tab:welfare} summarizes the cumulative welfare levels relative to the no-policy benchmark. In our setting, all transitions result in a welfare loss compared to the no-policy baseline. Achieving net zero emissions requires factoring in previously ignored external costs. This increases the relative price, directly or indirectly, of using the fossil resource both in housing and in energy production. The cost-effective way to achieve this is to impose a uniform carbon price on both housing and energy production. When comparing different constrained policy scenarios to the first best, the scenarios can be ranked according to their welfare impact. While the welfare is on lower levels in both second-best transitions, the phased-in carbon price transition leads to a higher welfare level than the subsidy scenario.

From a political economy perspective, transition costs are relevant. The transition creates direct costs for households by increasing both housing and energy costs, making energy demand more costly. We compare the cumulative housing and energy costs, during the transition. These costs do not correspond to the ranking in welfare. Although a uniform carbon price is the most cost-effective option, it creates substantial direct costs for households. Total housing expenses rise by 5 percent compared to the no-policy case. For the phased-in carbon price, the rise in housing costs is of a similar magnitude. In contrast, if the transition relies on subsidies for efficiency capital, the direct costs of households are lower than in the no policy case. The underlying driver is the difference in cumulative energy costs. While in both carbon pricing scenarios utility bills are 60 percent higher than in the no-policy benchmark, they are 25 percent lower in the subsidy-based transition. Hence, the least preferable option from a welfare perspective leads to the smallest direct costs of all scenarios. The underlying reason is that the energy demand is drastically reduced, which has a significant impact on housing costs. 

Redistributing carbon pricing revenues is an effective tool to lower the burden on households. The source of transfers are revenues from carbon pricing, both industry and housing. As the revenues from carbon pricing from the industry are identical in all scenarios, we focus only on the revenues from carbon pricing in housing. Thereby, we can measure the fiscal burden of the subsidy scenario. Subsidizing efficiency capital is costly and the government has to use lump-sum taxes to finance the subsidies. In the carbon price scenarios, households receive transfers that are equal to 1.5 percent of their cumulative income. At the same time, in the subsidy scenario, households have to pay taxes equal to 0.3 percent of their income. Consequently, households have a lower disposable income in the subsidy-based transition.This indicates how costly it is to decarbonize the economy through investment subsidies when the irreversibility constraint is binding. The need to invest in efficiency capital is largest in the first period, where revenues of carbon pricing in energy production are at their lowest level. Consequently, the government has to impose high taxes on households to finance the subsidies that are necessary to incentivize the large increase in efficiency capital due to the binding irreversibility constraint.

Relying on investment subsidies has important implications for housing demand. In detail, we find a housing-related rebound effect in the subsidy-based transition.\footnote{A graphical comparison of housing capital stocks for all scenarios is included in appendix \ref{app_fig}.} While the housing stock declines compared to the no policy benchmark, it is significantly higher than in the carbon price scenarios. This indicates that due to higher aggregate energy efficiency, households increase their housing consumption relative to the social optimum. Thereby, instrument choice not only matters for climate policy, but potentially affects also aggregate housing demand.

\section{Conclusion}

Due to a lack of investments, the building sector could become a barrier to the transition to a carbon-free economy.  Investment needs include the construction of new, energy-efficient buildings, along with the retrofitting of existing housing stock. In addition to housing-related investments, the green transition requires investments in renewable energy. Climate policy must incentivize and coordinate these various investments while recognizing the direct burden on households. In this nexus, we compared different transitions based on the availability of specific instruments in residential housing. We rely on a general equilibrium model with an elaborate setup of housing, energy production, and an optimizing government that chooses policy instruments optimally, given a set of constraints.

Climate policy has to not only incentivize these investments, but also coordinate between the different types of investments. Additionally, climate policy in housing can be difficult to implement, as it creates direct costs for households. In this nexus, we compared different transitions based on the availability of specific instruments in housing. We rely on a general equilibrium model with an elaborate setup of housing, energy production, and an optimizing government that chooses policy instruments optimally, given a set of constraints. 

If there are no constraints on mitigation policies, the conventional wisdom holds. The first-best policy is to impose a uniform carbon price on both households and the industry. Despite differences in housing investments, optimal climate policy remains unchanged. Energy demand only plays a secondary role in the optimal transition, as renewable energy expansion is not restricted in our setting. When comparing second-best transitions, the situation changes. While energy demand is not problematic with phased-in carbon prices, this is no longer the case in the subsidy-led transition. Due to the binding irreversibility constraint, the only option is to heavily subsidize investments in energy efficiency to reduce fossil fuel consumption. While providing subsidies for retrofits results in the lowest direct costs for households, it ultimately leads to the highest aggregate costs. Therefore, subsidies to efficiency investments prove to be an ineffective way to decarbonize the economy. 

We leave several important aspects for future research. This paper is on the role of energy demand, but other investments can additionally affect the substitution possibilities. For example, expanding infrastructure that enables district heating, especially in urban areas.  Furthermore, the housing sector is highly heterogeneous and is characterized by many barriers and additional market failures, including principal-agent problems and behavioral failures. This could be an promising aspect to consider, since the present analysis relies on perfect commitment and perfect foresight and thus anticipation of climate policy. In addition, market failures may limit the expansion of renewable energy and the amount of retrofits per year. Although we have an understanding of these obstacles in isolation, studying them in a general equilibrium context is essential for effective climate policy. A better understanding of their interactions will enhance climate policy's ability to navigate the complex task of decarbonizing the housing sector.

\newpage

\begin{singlespace}
\bibliographystyle{elsarticle-harv}
\bibliography{housing}

\end{singlespace}
\newpage

\appendix
\section{Analytical Derivations}
\label{app_analytical}
\subsection{No-Arbitrage Conditions}

Since the household's energy composite is given by
\begin{equation}
ene_{t}=\left(a_{ene}e_{t}^{\frac{\sigma_{ene}-1}{ene}}\left(1-a_{ene}\right)res_{t}^{\frac{\sigma_{ene}-1}{ene}}\right)^{\frac{\sigma_{ene}-1}{ene}}
\end{equation}
one can readily verify that the composite price of $ene_{t}$ equals
\begin{align}
p_{ene,t}= & \left(a_{ene}^{\sigma_{ene}}p_{E,t}^{1-\sigma_{ene}}+\left(1-a_{ene}\right)^{\sigma_{ene}}\left(p_{R,t}+p_{HC,t}\right)^{1-\sigma_{ene}}\right)^{\frac{1}{1-\sigma_{ene}}}
\end{align}

Using the consumer's first-order conditions lead to the following non-arbitrage
conditions

\begin{align}
R_{t}-\delta_{K}= & \left[MRS_{k_{t}^{H},c_{t}}-\delta_{H}-p_{ene,t}\kappa^{N}\right]+\frac{\phi_{t-1}^{H}}{\lambda_{t}}\left(1+\rho\right)-\frac{\phi_{t}^{H}}{\lambda_{t}}\left(1-\delta_{H}\right)\\
R_{t}-\delta_{K}= & \frac{\left(1-\tau_{INVE,t}\right)-\left(1-\tau_{INVE,t-1}\right)}{1-\tau_{INVE,t-1}}+\\
 & +\frac{\left[\frac{p_{ene,t}\overline{\kappa}}{k_{t}^{E}}\frac{\bar{k}}{k_{t}^{E}}-\left(1-\tau_{INVE,t}\right)\delta_{E}\right]+\frac{\phi_{t-1}^{E}}{\lambda_t}\left(1+\rho\right)-\frac{\phi_{t}^{E}}{\lambda_t}\left(1-\delta_{E}\right)}{1-\tau_{INVE,t}}\nonumber 
\end{align}

where $MRS_{k_{t}^{H},c_{t}}$ is the marginal rate of substitution between the consumer's housing capital  and the final good, $\tau_{INVE,t}$ is the subsidy to energy efficiency capital investment, $\lambda_t$ is the Lagrange multiplier associated with the consumer's budget constraint, and $\phi^H_t$ and $\phi^E_t$ are the multiplies associated with the non-negativity of investments.

\newpage

\subsection{Decentralized Setup}

The representative household-investor maximizes:
 \begin{equation}
     \max \sum_{t=0}^T \beta^t u(c_t,h_t)
 \end{equation}
 subject to:
 
\begin{align}
c_{t}+i_{t}^{Y}+i_{t}^{F}+i_{t}^{N}+i_{t}^{H}+i_{t}^{E}+\delta_{H}\overline{k}+p_{E,t}e_{t}+p_{R,t}res_{t} &  &  &  & \left(\lambda_{t}\right)\label{eq:BC}\\
=w_{t}l+R_{t}k_{t}+R_{t}^{F}k_{t}^{F}+R_{t}^{N}k_{t}^{N}\nonumber 
\end{align}

\begin{subequations}
\begin{align}
ene_{t}= & h^{e}\left(e_{t},res_{t}\right)=\left(a_{ene}e_{t}^{\frac{\sigma_{ene}-1}{ene}}+\left(1-a_{ene}\right)res_{t}^{\frac{\sigma_{ene}-1}{ene}}\right)^{\frac{\sigma_{ene}-1}{ene}} & \left(\nu_{t}^{e}\right)\label{eq:ene}\\
ene_{t}= & h^{ene}\left(k_{t}^{E},k_{t}^{H}\right)=\left(\kappa_{N}+\frac{\overline{\kappa}}{k_{t}^{E}}\right)\overline{k}+\kappa_{N}k_{t}^{H} & \left(\nu_{t}^{ene}\right)\\
h_{t}= & h^{k}\left(land_{t},k_{t}^{H}\right)=land_{t}^{a_{h}}\left(\overline{k}+k_{t}^{H}\right)^{1-a_{h}} & \left(\chi_{t}\right)\\
k_{t+1}^{Y}= & i_{t}^{Y}+\left(1-\delta_{Y}\right)k_{t}^{Y} & \left(\psi_{t}\right)\\
k_{t+1}^{F}= & i_{t}^{F}+\left(1-\delta_{F}\right)k_{t}^{F} & \left(\psi_{t}^{F}\right)\\
k_{t+1}^{N}= & i_{t}^{N}+\left(1-\delta_{H}\right)k_{t}^{N} & \left(\psi_{t}^{N}\right)\\
k_{t+1}^{H}= & i_{t}^{H}+\left(1-\delta_{F}\right)k_{t}^{H} & \left(\psi_{t}^{H}\right)\\
k_{t+1}^{E}= & i_{t}^{E}+\left(1-\delta_{N}\right)k_{t}^{E} & \left(\psi_{t}^{E}\right)\\
i_{t}^{H}\geq & 0 & \left(\phi_{t}^{H}\right)\\
i_{t}^{E}\geq & 0 & \left(\phi_{t}^{E}\right)\label{eq:iEPositive}
\end{align}
\end{subequations}

The first-order conditions are as follows:

\begin{subequations}
\begin{align}
\frac{\partial\mathcal{L}}{\partial c_{t}}= & u_{c_{t}}-\lambda_{t}=0 & \Rightarrow & \lambda_{t}=u_{c_{t}}\\
\frac{\partial\mathcal{L}}{\partial h_{t}}= & u_{h_{t}}-\chi_{t}=0 & \Rightarrow & u_{h_{t}}=\chi_{t}\\
\frac{\partial\mathcal{L}}{\partial ene_{t}}= & -\nu_{t}^{e}+\nu_{t}^{ene}=0 & \Rightarrow & \nu_{t}^{e}=\nu_{t}^{ene}\\
\frac{\partial\mathcal{L}}{\partial e{}_{t}}= & \nu_{t}^{e}\frac{\partial h^{e}}{\partial e{}_{t}}-\lambda_{t}p_{E,t}=0 & \Rightarrow & \lambda_{t}p_{E,t}=\nu_{t}^{e}\frac{\partial h^{e}}{\partial e{}_{t}}\\
\frac{\partial\mathcal{L}}{\partial res{}_{t}}= & \nu_{t}^{e}\frac{\partial h^{e}}{\partial res{}_{t}}-\lambda_{t}p_{R,t}=0 & \Rightarrow & \lambda_{t}p_{R,t}=\nu_{t}^{e}\frac{\partial h^{e}}{\partial res{}_{t}}\\
\frac{\partial\mathcal{L}}{\partial i_{t}^{Y}}= & -\lambda_{t}+\psi_{t}=0 & \Rightarrow & \lambda_{t}=\psi_{t}\\
\frac{\partial\mathcal{L}}{\partial i_{t}^{F}}= & -\lambda_{t}+\psi_{t}^{F}=0 & \Rightarrow & \lambda_{t}=\psi_{t}^{F}\\
\frac{\partial\mathcal{L}}{\partial i_{t}^{N}}= & -\lambda_{t}+\psi_{t}^{N}=0 & \Rightarrow & \lambda_{t}=\psi_{t}^{N}\\
\frac{\partial\mathcal{L}}{\partial i_{t}^{H}}= & -\lambda_{t}+\psi_{t}^{H}+\phi_{t}^{H}=0 & \Rightarrow & \lambda_{t}=\psi_{t}^{H}+\phi_{t}^{H}\\
\frac{\partial\mathcal{L}}{\partial i_{t}^{E}}= & -\lambda_{t}+\psi_{t}^{E}+\phi_{t}^{E}=0 & \Rightarrow & \lambda_{t}=\psi_{t}^{E}+\phi_{t}^{E}
\end{align}
\end{subequations}

\begin{align}
\frac{\partial\mathcal{L}}{\partial k_{t+1}^{Y}}= & -\beta^{t}\psi_{t}+\beta^{t+1}\lambda_{t+1}R_{t+1}+\left(1-\delta_{Y}\right)\beta^{t+1}\psi_{t+1}=0\label{eq:FOCk}\\
\Rightarrow R_{t+1}= & \frac{1}{\lambda_{t+1}}\left[\psi_{t}/\beta-\left(1-\delta_{Y}\right)\psi_{t+1}\right]\\
= & \frac{\lambda_{t}}{\beta\lambda_{t+1}}-\left(1-\delta_{Y}\right)\\
= & \frac{u_{c_{t}}}{\beta u_{c_{t+1}}}-\left(1-\delta_{Y}\right)
\end{align}

\begin{align}
\frac{\partial\mathcal{L}}{\partial k_{t+1}^{j}}= & -\beta^{t}\psi_{t}^{j}+\beta^{t+1}\lambda_{t+1}R_{t+1}^{j}+\left(1-\delta_{j}\right)\beta^{t+1}\psi_{t+1}^{j}=0 & \text{for \ensuremath{j=F,N}}\label{eq:FOCkFN}\\
\Rightarrow R_{t+1}^{j}=\frac{1}{\lambda_{t+1}} & \left[\psi_{t}^{j}/\beta-\left(1-\delta_{j}\right)\psi_{t+1}^{j}\right]=R_{t+1}\text{ \text{for \ensuremath{j=F,N}}}
\end{align}

\begin{align}
\frac{\partial\mathcal{L}}{\partial k_{t+1}^{H}}= & -\beta^{t}\psi_{t}^{H}+\beta^{t+1}\left[-\nu_{t+1}^{ene}\frac{\partial h^{ene}}{\partial k_{t+1}^{H}}+\chi_{t+1}\frac{\partial h^{k}}{\partial k_{t+1}^{H}}+\left(1-\delta_{H}\right)\psi_{t+1}^{H}\right]=0\label{eq:FOCkH}\\
\Rightarrow\frac{\partial h^{k}}{\partial k_{t+1}^{H}}= & \frac{1}{\chi_{t+1}}\left[\psi_{t}^{H}/\beta+\nu_{t+1}^{ene}\frac{\partial h^{ene}}{\partial k_{t+1}^{H}}-\left(1-\delta_{H}\right)\psi_{t+1}^{H}\right]\\
\Rightarrow\frac{\partial h^{ene}}{\partial k_{t+1}^{H}}= & \frac{1}{\nu_{t+1}^{ene}}\left[\chi_{t+1}\frac{\partial h^{k}}{\partial k_{t+1}^{H}}+\left(1-\delta_{H}\right)\psi_{t+1}^{H}-\psi_{t}^{H}/\beta\right]
\end{align}

\begin{align}
\frac{\partial\mathcal{L}}{\partial k_{t+1}^{E}}= & -\beta^{t}\psi_{t}^{E}+\beta^{t+1}\left[-\nu_{t+1}^{ene}\frac{\partial h^{ene}}{\partial k_{t+1}^{E}}+\left(1-\delta_{E}\right)\psi_{t+1}^{E}\right]=0\label{eq:FOCkE}\\
\Rightarrow\frac{\partial h^{ene}}{\partial k_{t+1}^{E}}= & \frac{1}{\nu_{t+1}^{ene}}\left[\left(1-\delta_{E}\right)\psi_{t+1}^{E}-\psi_{t}^{E}/\beta\right]
\end{align}

First, we can recognize the implicit rental rates of the different types of capital in spirit of \citet{jorgenson1967theory}. The implicit rental rate of physical capital $k_t$ is defined as:

\begin{equation}
R_{t}=\frac{\psi_{t-1}}{\beta\psi_{t}}-\left(1-\delta_{Y}\right)
\end{equation}

where $\psi_t$ is the shadow price of physical capital. As capital freely moves between final good production and fossil and clean energy production, their rental rates are equalized. That is

\begin{equation}
R_{t}=R^F_t=R^N_t
\end{equation} 

\subsection{Optimal Carbon Price}
 The aim of this section is to derive some analytical results. In what follows we write the emissions budget is a recursive form. Since the economy consists of $n$ identical households, the social planner problem is as follows:

 \begin{equation}
     \max \sum_{t=0}^T \beta^t n u(c_t,h_t)
 \end{equation}

 subject to equations $\left(\ref{eq:ene}\right)-\left(\ref{eq:iEPositive}\right)$ multiplied by the number of households $n$ and also subject to

\begin{align}
F^{Y}\left(nk_{t}^{Y},E_{Y,t}\right)= & n\left(c_{t}+i_{t}^{Y}+i_{t}^{F}+i_{t}^{N}+i_{t}^{H}+i_{t}^{E}+\delta_{H}\overline{k}\right) & \left(\lambda_{t}\right)\label{eq:Acct}\\
 & +p_{R,t}\left(n\times res_{t}+Res_{F,t}\right)\nonumber \\
E_{t}=F^{E}\left(E_{F,t},E_{N,t}\right)= & \left(a_{E}E_{F,t}^{\frac{\sigma_{E}-1}{\sigma_{E}}}+\left(1-a_{E}\right)E_{N,t}^{\frac{\sigma_{E}-1}{\sigma_{E}}}\right)^{\frac{\sigma_{E}}{\sigma_{E}-1}} & \left(\chi_{t}^{Y}\right)\\
E_{F,t}=F^{F}\left(nk_{t}^{F},Res_{F,t}\right)= & \left(a_{F}\left(nk_{t}^{F}\right)^{\frac{\sigma_{F}-1}{\sigma_{F}}}+\left(1-a_{F}\right)Res_{F,t}^{\frac{\sigma_{F}-1}{\sigma_{F}}}\right)^{\frac{\sigma_{F}}{\sigma_{F}-1}} & \left(\chi_{t}^{F}\right)\\
E_{N,t}=F^{N}\left(nk_{t}^{N}\right)= & A_{N,t}nk_{t}^{N} & \left(\chi_{t}^{N}\right)\\
M_{t+1}= & M_{t}-\left(n\times res_{t}+Res_{F,t}\right) & \left(\mu_{t}^{R}\right)\\
M_{t}\geq & 0 & \left(\mu_{t}\right)\\
E_{t}= & n\times e_{t}+E_{Y,t} & \left(\chi_{t}^{E}\right)\label{eq:FCE}
\end{align}

Note that the variable in parenthesis to the right denotes the Lagrangian multiplier associated with the equation at hand. The first order conditions of the social planner problem are as follows:

\begin{subequations}
\begin{align}
\frac{\partial\mathcal{L}}{\partial c_{t}}= & nu_{c_{t}}-n\lambda_{t}=0 & \Rightarrow & \lambda_{t}=u_{c_{t}}\\
\frac{\partial\mathcal{L}}{\partial h_{t}}= & nu_{h_{t}}-n\chi_{t}=0 & \Rightarrow & \chi_{t}=u_{h_{t}}\\
\frac{\partial\mathcal{L}}{\partial ene_{t}}= & -n\nu_{t}^{e}+n\nu_{t}^{ene}=0 & \Rightarrow & \nu_{t}^{e}=\nu_{t}^{ene}\\
\frac{\partial\mathcal{L}}{\partial e{}_{t}}= & n\nu_{t}^{e}\frac{\partial h^{e}}{\partial e_{t}}-n\chi_{t}^{E}=0 & \Rightarrow & \chi_{t}^{E}=\nu_{t}^{e}\frac{\partial h^{e}}{\partial e_{t}}\\
\frac{\partial\mathcal{L}}{\partial res_{t}}= & -n\lambda_{t}p_{R,t}+n\nu_{t}^{e}\frac{\partial h^{e}}{\partial res_{t}}-n\mu_{t}^{R}=0 & \Rightarrow & \mu_{t}^{R}+\lambda_{t}p_{R,t}=\nu_{t}^{e}\frac{\partial h^{e}}{\partial res_{t}}\\
\frac{\partial\mathcal{L}}{\partial i_{t}^{Y}}= & -n\lambda_{t}+n\psi_{t}=0 & \Rightarrow & \lambda_{t}=\psi_{t}\\
\frac{\partial\mathcal{L}}{\partial i_{t}^{F}}= & -n\lambda_{t}+n\psi_{t}^{F}=0 & \Rightarrow & \lambda_{t}=\psi_{t}^{F}\\
\frac{\partial\mathcal{L}}{\partial i_{t}^{N}}= & -n\lambda_{t}+n\psi_{t}^{N}=0 & \Rightarrow & \lambda_{t}=\psi_{t}^{N}\\
\frac{\partial\mathcal{L}}{\partial i_{t}^{H}}= & n\left(-\lambda_{t}+\psi_{t}^{H}+\phi_{t}^{H}\right)=0 & \Rightarrow & \lambda_{t}=\psi_{t}^{H}+\phi_{t}^{H}\\
\frac{\partial\mathcal{L}}{\partial i_{t}^{E}}= & n\left(-\lambda_{t}+\psi_{t}^{E}+\phi_{t}^{E}\right)=0 & \Rightarrow & \lambda_{t}=\psi_{t}^{E}+\phi_{t}^{E}
\end{align}
\end{subequations}

\begin{align}
\frac{\partial\mathcal{L}}{\partial E_{Y,t}}= & \lambda_{t}\frac{\partial F^{Y}}{\partial E_{Y,t}}-\chi_{t}^{E}=0 & \Rightarrow & \chi_{t}^{E}=\lambda_{t}\frac{\partial F^{Y}}{\partial E_{Y,t}}\\
\frac{\partial\mathcal{L}}{\partial E_{t}}= & -\chi_{t}^{Y}+\chi_{t}^{E}=0 & \Rightarrow & \chi_{t}^{Y}=\chi_{t}^{E}\\
\frac{\partial\mathcal{L}}{\partial E_{F,t}}= & \chi_{t}^{Y}\frac{\partial F^{E}}{\partial E_{F,t}}-\chi_{t}^{F}=0 & \Rightarrow & \chi_{t}^{F}=\chi_{t}^{Y}\frac{\partial F^{E}}{\partial E_{F,t}}\\
\frac{\partial\mathcal{L}}{\partial E_{N,t}}= & \chi_{t}^{Y}\frac{\partial F^{E}}{\partial E_{N,t}}-\chi_{t}^{N}=0 & \Rightarrow & \chi_{t}^{N}=\chi_{t}^{Y}\frac{\partial F^{E}}{\partial E_{N,t}}\\
\frac{\partial\mathcal{L}}{\partial Res_{F,t}}= & -\lambda_{t}p_{R,t}+\chi_{t}^{F}\frac{\partial F^{F}}{\partial Res_{F,t}}-\mu_{t}^{R}=0 & \Rightarrow & \lambda_{t}p_{R,t}+\mu_{t}^{R}=\chi_{t}^{F}\frac{\partial F^{F}}{\partial Res_{F,t}}
\end{align}

\begin{align}
\frac{\partial\mathcal{L}}{\partial k_{t+1}^{Y}}= & 0\label{eq:FOCSPkY-1}\\
\Rightarrow\frac{\partial F^{Y}}{\partial nk_{t+1}^{Y}}= & \frac{1}{\lambda_{t+1}}\left[\psi_{t}/\beta-\left(1-\delta_{Y}\right)\psi_{t+1}\right]
\end{align}

\begin{align}
\frac{\partial\mathcal{L}}{\partial k_{t+1}^{F}}= & 0\label{eq:FOCSPkF-1}\\
\Rightarrow\frac{\partial F^{F}}{\partial nk_{t+1}^{F}}= & \frac{1}{\chi_{t}^{F}}\left[\psi_{t}^{F}/\beta-\left(1-\delta_{F}\right)\psi_{t+1}^{F}\right]
\end{align}

\begin{align}
\frac{\partial\mathcal{L}}{\partial k_{t+1}^{N}}= & 0\label{eq:FOCkN-1}\\
\Rightarrow\frac{\partial F^{N}}{\partial nk_{t+1}^{N}}= & \frac{1}{\chi_{t+1}^{N}}\left[\psi_{t}^{N}/\beta-\left(1-\delta_{N}\right)\psi_{t+1}^{N}\right]
\end{align}

\begin{align}
\frac{\partial\mathcal{L}}{\partial k_{t+1}^{H}} & =0\label{eq:FOCSPkH-1}\\
\Rightarrow\frac{\partial h^{k}}{\partial k_{t+1}^{H}}= & \frac{1}{\chi_{t+1}}\left[\psi_{t}^{H}/\beta+\nu_{t+1}^{ene}\frac{\partial h^{ene}}{\partial k_{t+1}^{H}}-\left(1-\delta_{H}\right)\psi_{t+1}^{H}\right]\\
\frac{\partial h^{ene}}{\partial k_{t+1}^{H}}= & \frac{1}{\nu_{t+1}^{ene}}\left[\chi_{t+1}\frac{\partial h^{k}}{\partial k_{t+1}^{H}}+\left(1-\delta_{H}\right)\psi_{t+1}^{H}-\psi_{t}^{H}/\beta\right]
\end{align}

\begin{align}
\frac{\partial\mathcal{L}}{\partial k_{t+1}^{E}}= & 0\label{eq:FOCSPkE-1}\\
\Rightarrow\frac{\partial h^{ene}}{\partial k_{t+1}^{E}}= & \frac{1}{\nu_{t+1}^{ene}}\left[\left(1-\delta_{H}\right)\psi_{t+1}^{H}-\psi_{t}^{E}/\beta\right]=0
\end{align}

\begin{align}
\frac{\partial\mathcal{L}}{\partial M_{t+1}}= & -\beta^{t}\mu_{t}^{R}+\beta^{t+1}\mu_{t+1}^{R}+\beta^{t+1}\mu_{t+1}=0\label{eq:FOCSPM-1}\\
\Rightarrow\mu_{t+1}^{R}+\mu_{t+1} & =\frac{\mu_{t}^{R}}{\beta}
\end{align}

Since the marginal product of $Res_{t}$ approaches infinity as $Res_{t}$
approaches zero and since $res_{t}$ must be used for housing services
then $M_{t}$ approaches zero only at the end of the horizon $t^{*}$.
Implying that $\mu_{t}=0$ for all $t<t^{*}.$ Therefore

\begin{equation}
\frac{\mu_{t+1}^{R}-\mu_{t}^{R}}{\mu_{t}^{R}}=\frac{1-\beta}{\beta}
\end{equation}

The aim of this subsection is to analyze optimal carbon taxation. The underlying question is whether fossil resource use in housing and in energy production should be taxed in the same way. Let us start with the optimal carbon tax for households. Their optimal resource use is described by:

\[
\lambda_t p_{R,t} (1+\tau^H_t) = \nu_t^e \frac{\partial h^e(e_t,res_t)}{\partial res_t} 
\]
\[
(1+\tau^H_t) = \frac{\nu_t^e }{\lambda_t p_{R,t}} \frac{\partial h^e(e_t,res_t)}{\partial res_t}
= \frac{\nu_t^e }{u_{c_{t}} p_{R,t}}\frac{\partial h^e(e_t,res_t)}{\partial res_t}
\]

A carbon price can be implemented instead of a carbon tax, by substituting $p_{R,t}\tau^H_t$ with $p_{HC,t}$.  Recall that from Social planner optimization, we know: 
\[
\nu_t^e \frac{\partial h^e(e_t,res_t)}{\partial res_t} = \mu_t^R + \lambda_t p_{R,t} = \mu_t^R + u_{c_t} p_{R,t}
\]

If we plug it in, we get: 
\[
(1+\tau_t^H) = \frac{\nu_t^e }{\lambda_t p_{R,t}} \frac{\partial h^e(e_t,res_t)}{\partial res_t}
= \frac{\mu_t^R + u_{c_t} p_{R,t}}{u_{c_t} p_{R,t}}
\]
Thus, we arrive at the optimal carbon tax of households:
\begin{equation}
    \tau_t^H=\frac{\mu_t^R}{u_{c_t} p_{R,t}}
\end{equation}

Next, we can study the optimal carbon tax in fossil energy production. The profit maximization problem of the fossil energy producer is:
\begin{equation}
    \pi_{F,t}= p_{F,t} F^F(K_{F,t},Res_{F,t}) - R^F_t K_{F,t} - p_{R,t} (1+\tau_t^Y) Res_{F,t}.
\end{equation}
The first-order condition with regard to capital equals
\begin{equation}
    p_{F,t} \frac{\partial F^F(K_{F,t},Res_{F,t})}{\partial K_{F,t}}  = R^F_t.
\end{equation}
The optimal resource use for fossil energy producers is described by:

\begin{equation}
    p_{F,t} \frac{\partial F^F(K_{F,t},Res_{F,t})}{\partial Res_{F,t}} = p_{R,t} (1+\tau_t^Y)
\end{equation}

\[
 (1+\tau_t^Y) \frac{p_{R,t}}{p_{F,t}}= F^F_{Res} 
\]
From social planner optimization, we use:
\[
F^F_{Res}=\frac{\mu^R_t}{\chi^F_t}+\frac{u_{c_t} p_{R,t}}{\chi^F_t}
\]
If we plug it in the optimal resource use condition, we get:
\[
 (1+\tau_t^Y) \frac{p_{R,t}}{p_{F,t}}= \frac{\mu^R_t}{\chi^F_t}+\frac{u_{c_t} p_{R,t}}{\chi^F_t}
\]

or

\[
 (1+\tau_t^Y) = \frac{\mu_t^R p_{F,t}}{\chi^F_t p_{R,t}}+\frac{u_{c_t} p_{F,t}}{\chi^F_t}
\]
we arrive at:
\begin{equation}
    \tau_t^Y = \frac{\mu_t^R p_{F,t}}{\chi^F_t p_{R,t}}+\frac{u_{c_t} p_{F,t}}{\chi^F_t}-1
\end{equation}

The open question is how the shadow price of fossil energy $\chi^F_t$ relates to the multiplier of the resource constraint $\lambda_t$ and thus to the marginal utility $u_{c_t}$. Recall that the shadow price of fossil energy is:
\[
\chi^F_t= \chi^Y_t F_{E_{F,t}}^F = \chi^E_t F_{E_{F,t}}^F = \lambda_t \frac{\partial  F^Y}{\partial E_{Y,t}} F_{E_{F,t}}^F
\]

We now use expressions from the profit maximization of final energy production and the final good producer to simply. Recall that the final energy producer maximizes:

\begin{equation}
    \pi_{E,t} = p_{E,t} F^E(E_{F,t},E_{N,t}) - p_{F,t} E_{F,t} - p_{N,t} E_{N,t}
\end{equation}

with the respective first order conditions:

\begin{equation}
    p_{F,t} = \frac{\partial F^E(E_{F,t},E_{N,t})}{\partial E_{F,t}} p_{E,t}
\end{equation}

\begin{equation}
    p_{N,t} = \frac{\partial F^E(E_{F,t},E_{N,t})}{\partial E_{N,t}} p_{E,t}
\end{equation}
From the profit maximization of the final good producer we can get an expression for the price of final energy $p_t^E$:
\begin{equation}
p_{E,t} = \frac{\partial F^Y(K_{Y,t},E_{Y,t})}{\partial E_{Y,t}}     
\end{equation}

Now, we can combine both to get:
\begin{equation}
    p_{F,t} =  \frac{\partial F^Y(K_{Y,t},E_{Y,t})}{\partial E_{Y,t}} \frac{\partial F^E(E_{F,t},E_{N,t})}{\partial E_{F,t}}
\end{equation}
we can substitute the shadow price of aggregate energy:
\[
\chi_t^F= \lambda_t p_{F,t}
\]

If we plug it into the formula for the optimal carbon taxes in energy production, we get:

\begin{equation}
    \tau_t^Y = \frac{\mu_t^R}{u_{c_t} p_{R,t}} = \tau_t^H
\end{equation}

Thus, the optimal carbon price for fossil energy producers is identical to the carbon price for households.

\section{Figures}
\label{app_fig}

\begin{figure}
    \centering
    \includegraphics[scale=0.85]{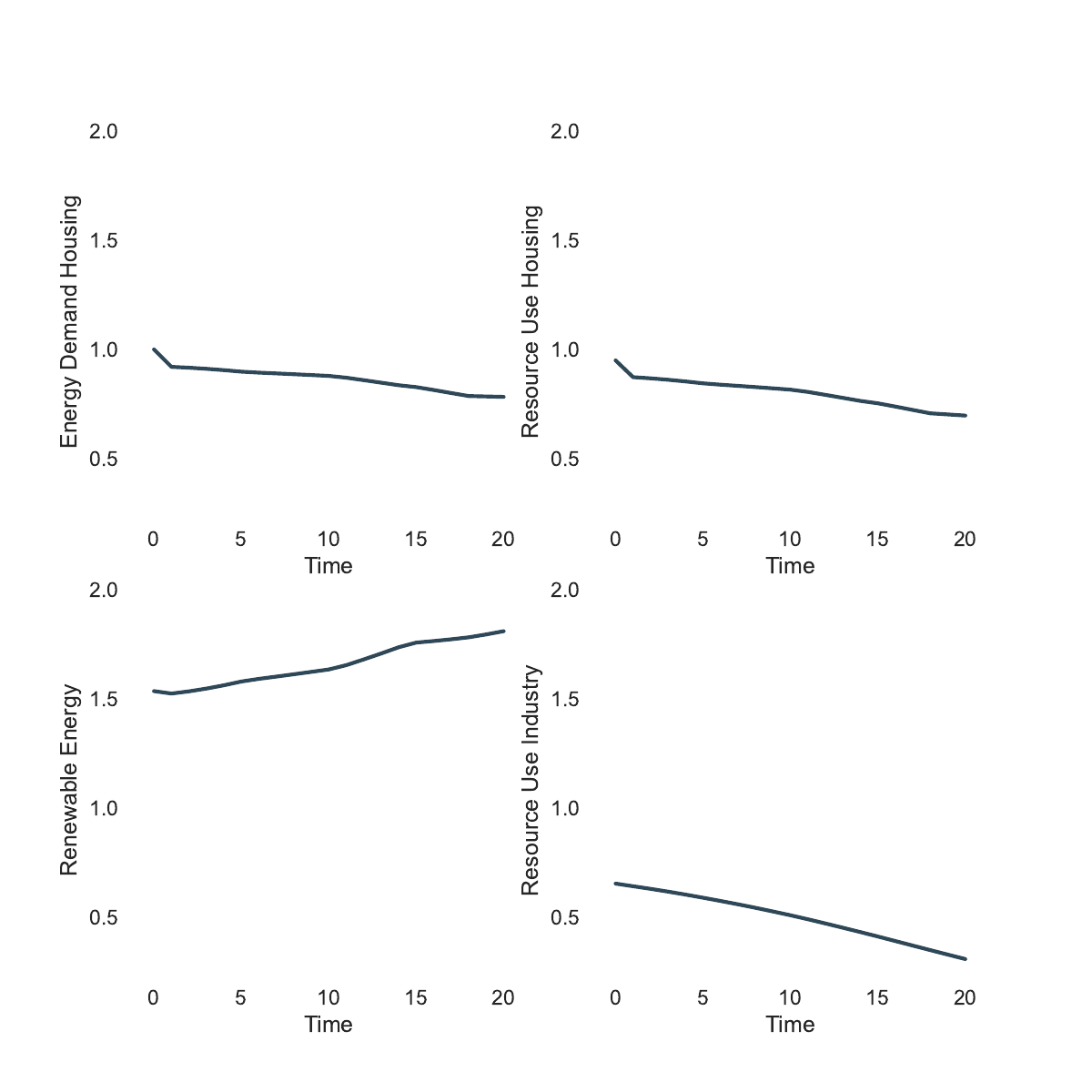}
    \caption{Evolution of housing-related energy demand, resource use in housing and industry and the expansion of renewable energy in the first best. Variables are expressed in deviations from the no-policy benchmark.}
    \label{fig:enter-label}
\end{figure}

\begin{figure}
    \centering
    \includegraphics[scale=0.8]{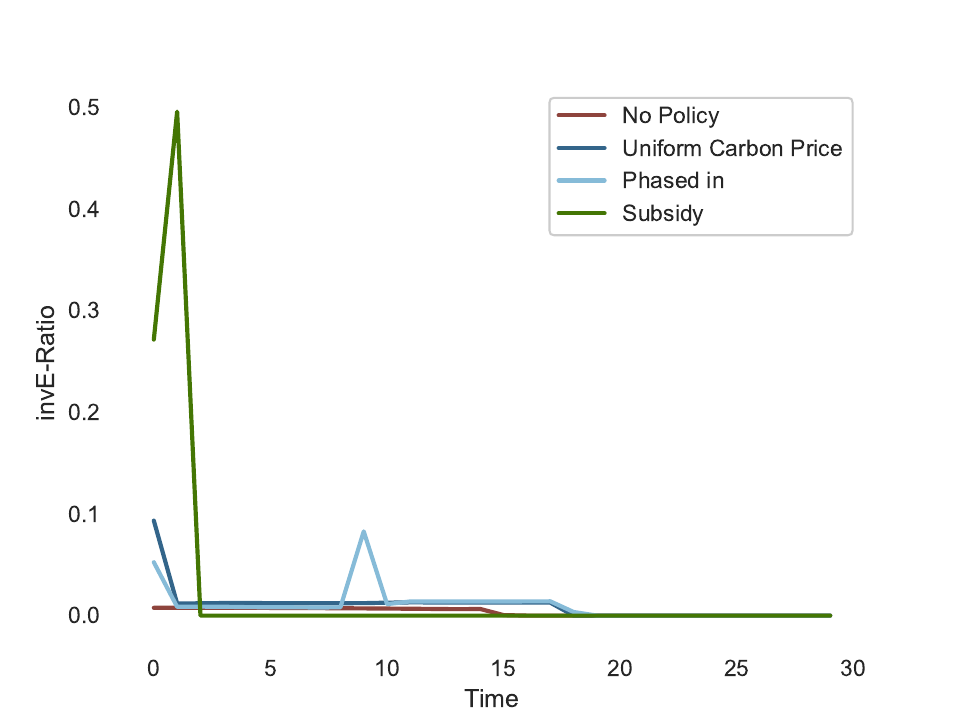}
    \caption{Efficiency capital investments in the optimal policy scenario, second-best scenarios and the no-policy baseline. Investments are expressed as a ratio over physical capital investments.}
    \label{invE}
\end{figure}

\begin{figure}
    \centering
    \includegraphics[scale=0.8]{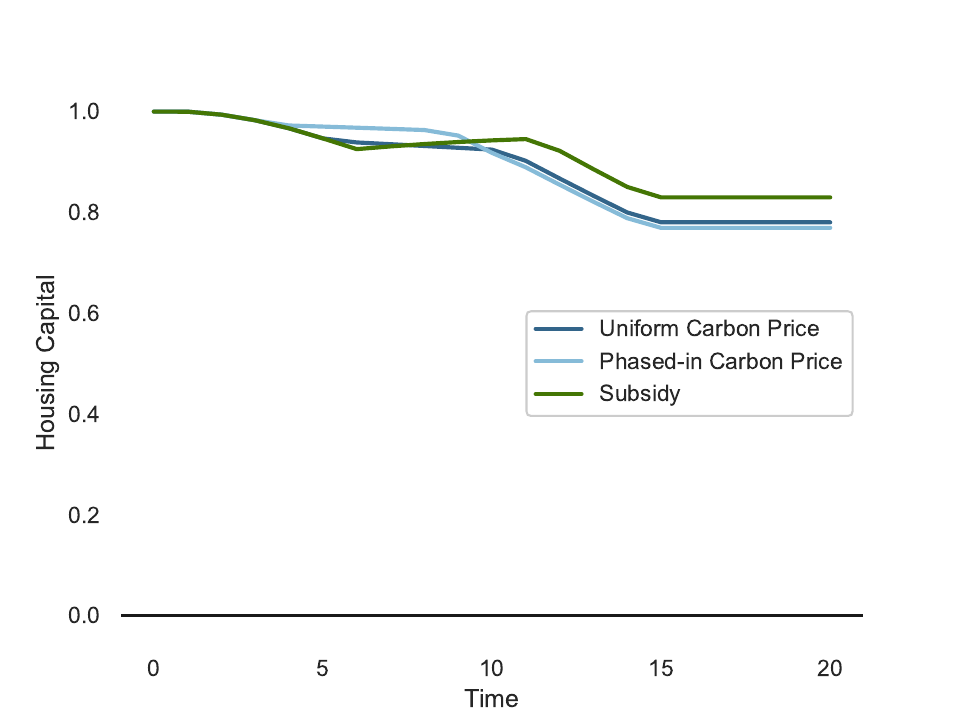}
    \caption{Housing capital stock in the optimal policy scenario, second-best scenarios and the no-policy baseline. Investments are expressed as a ratio over physical capital investments.}
    \label{khComp}
\end{figure}

\begin{figure}
    \centering
    \includegraphics[scale=0.8]{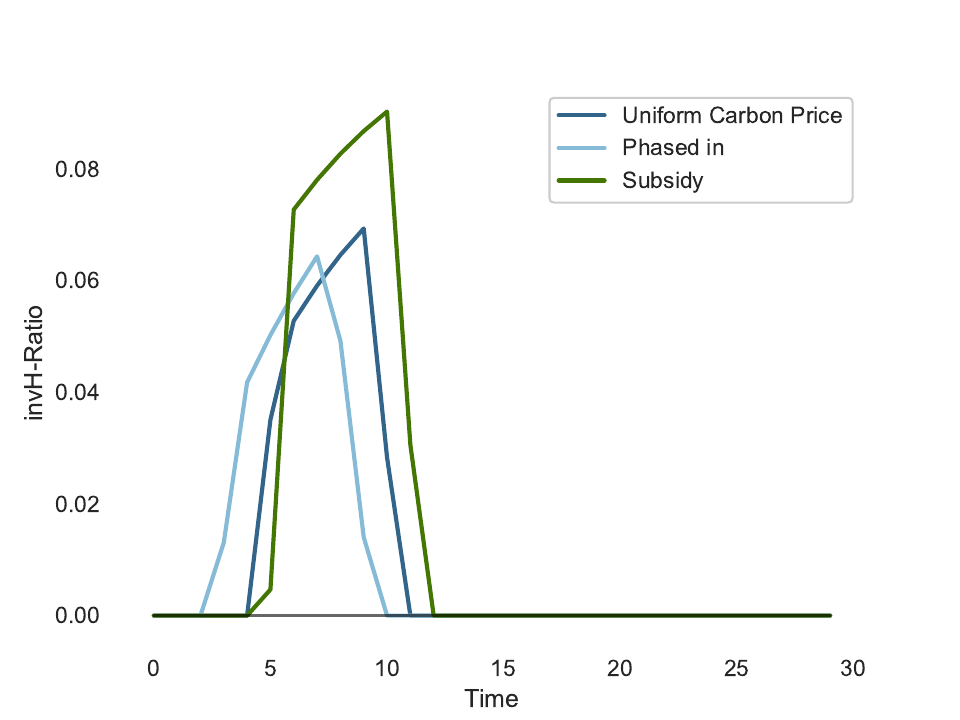}
    \caption{Housing capital investments in the optimal policy scenario, second-best scenarios and the no-policy baseline. Investments are expressed as a ratio over physical capital investments.}
    \label{invH2}
\end{figure}

\end{document}